\documentclass[epj]{svjour}

\usepackage{epsf}
\usepackage[T1]{fontenc}
\usepackage{amsmath}
\usepackage{amsfonts}
\usepackage{amssymb}
\usepackage{graphicx}
\usepackage{psfrag}

\newcommand{\beq}{\begin{equation}}
\newcommand{\eeq}{\end{equation}}
\newcommand{\beqn}{\begin{eqnarray}}
\newcommand{\eeqn}{\end{eqnarray}}

\begin{document}

%opening
\title{Klein tunneling in graphene: optics with massless electrons}
\titlerunning{Klein tunneling in graphene}
\author{Pierre E. Allain\inst{1} \and Jean-No\"el Fuchs\inst{2}\thanks{\email{jean-noel.fuchs@u-psud.fr}} }
\institute{\inst{1}{Institut d'\'Electronique Fondamentale, Univ. Paris-Sud,
CNRS UMR 8622, F-91405 Orsay, France} \\
\inst{2}{Laboratoire de Physique
des Solides, Univ. Paris-Sud, CNRS, UMR 8502, F-91405 Orsay, France}}
\date{\today}
\abstract{This article provides a pedagogical review on Klein tunneling in
graphene, i.e. the peculiar tunneling properties of two-dimensional
massless Dirac electrons. We consider two simple situations in
detail: a massless Dirac electron incident either on a potential
step or on a potential barrier and use elementary quantum wave
mechanics to obtain the transmission probability. We emphasize the
connection to related phenomena in optics, such as the
Snell-Descartes law of refraction, total internal reflection,
Fabry-P\'erot resonances, negative refraction index materials
(the so called meta-materials), etc. We also stress that Klein tunneling is not a
genuine quantum tunneling effect as it does not necessarily involve passing through
a classically forbidden region via evanescent waves. A crucial role in Klein tunneling
is played by the conservation of (sublattice) pseudo-spin, which is discussed in detail.
A major consequence is the absence of backscattering at normal incidence, of which we give a new shorten proof.
The current experimental status is also thoroughly reviewed. The appendix contains the discussion of
a one-dimensional toy model that clearly illustrates the difference in Klein tunneling
between mono- and bi-layer graphene.}
%\PACS{
%      {87.17.Jj}{Cell locomotion; chemotaxis} \and
%      {87.17.Rt}{Cell adhesion and cell mechanic}
%     }

\maketitle

\section{Introduction}
The tunnel effect of a particle going through a potential barrier is now
a standard exercise in elementary quantum mechanics, which goes back
to the early days of this theory \cite{Gamow1928,GurneyCondon1928}.
It is usually obtained by solving the Schr\"odinger equation either
approximately with the semiclassical WKB method or exactly for
piecewise constant (square) potentials \cite{Messiah}. The
probability for the particle to cross the potential barrier decays
exponentially with the width and the energy height of the barrier.
Thus, even if classically the probability to go through the barrier
is equal to zero, quantum dynamics allows the crossing with a tiny
probability. The mechanism behind quantum tunneling is the possibility
for a quantum particle to enter a classically forbidden regions as permitted
by Heisenberg's uncertainty principle. This is possible thanks to evanescent waves.

However it would be considered paradoxical for a
particle to tunnel with certainty regardless of the height and the
width of the barrier. It turns out that such an effect has been
described theoretically by the Swedish physicist Oskar Klein in 1929
\cite{klein,calogeracos} for relativistic electrons using the three
dimensional (3D) massive Dirac equation (i.e. the original Dirac
equation describing a relativistic massive electron \cite{Dirac} in
ordinary space). This effect has since been know as the ``Klein paradox''.

Recently a similar effect, though for 2D massless Dirac electrons,
has been predicted \cite{Katsnelson} (see also \cite{Cheianov} and
\cite{pereira}) and evidences of its observation in a graphene sheet
were reported \cite{huard,gorbachev,stander,kim}. The latter is now known as
the ``Klein tunnel effect'' or ``Klein tunneling''.

The aim of the present article is to review this effect and to
present simple derivations that can easily be reproduced in order to
demystify it. In particular, we wish to emphasize that the ``Klein
tunnel effect'' is \emph{not} a tunnel effect in the usual quantum
mechanical sense as it does not crucially rely on evanescent waves
and that it is not paradoxical (at least from the solid state physics
perspective). For example, there is no problem with charge conservation as often stated in the context of the Klein paradox.
We also give simple physical arguments in order to understand the unusual tunneling behavior of
massless Dirac particles and stress the crucial importance of the
(sublattice) pseudo-spin conservation.

The article is organized as follows. In section \ref{2dmdeig}, we
review the low-energy effective theory of valence electrons in
graphene and then discuss properties of the 2D massless Dirac
equations that are to be used in the following sections. In particular, we emphasize
the importance of the pseudo-spin conservation and prove that it implies the
absence of backscattering at normal incidence on any potential profile. Section
\ref{secpotstep} discusses the case of a massless Dirac electron
incident on a potential step (both sharp and smooth)
and section \ref{barrier} that of a square potential barrier.
In section \ref{exps}, we review in detail the current status of experiments
on Klein tunneling. Conclusions are given in section \ref{ccl}. An appendix treats
Klein tunneling in 1D using a toy model that allows us to easily compare
mono- and bi-layer graphene.

We mention the following reviews on the Klein paradox by
Calogeracos and Dombey \cite{calogeracos,calogeracos2} and more
recently on Klein tunneling in graphene by
Beenakker \cite{Beenakker2008} and by Pereira et al. \cite{Pereira}.

\section{Two-dimensional massless Dirac equation in
graphene}\label{2dmdeig}
\subsection{Low energy description of electrons in graphene}
%%%%%%%%%%%%%%%
%\begin{figure}[h]
%\centering
%\includegraphics[width=5cm]{honeycombstructure.pdf}
%\caption{Honeycomb crystal of graphene with the two triangular sublattices, where $\bf{a1}$ and $\bf{a2}$ are the lattice vectors of the triangular Bravais lattice. One sublattice is represented by the atom of type A the other sublattice with the atom of type B.}
%\label{honeycomb}
%\end{figure}
%%%%%%%%%%%%%%%
We start by briefly reviewing the low-energy effective description
of valence electrons in a graphene flake. Graphene is a two
dimensional honeycomb crystal of carbon atoms. The honeycomb crystal
is however not a Bravais lattice. It is made out of a triangular
Bravais lattice with a two atom basis (usually called $A$ and $B$).
This can alternately be seen as two triangular sublattices. As a
consequence of this two-site basis, the electronic wavefunction is a
bispinor: in other words, the electron carries -- in addition to its
usual spin $1/2$, which we shall neglect in the following -- a
pseudo-spin $1/2$ associated with its sublattice degree of freedom.
We shall refer to it as sublattice pseudo-spin. The electronic band
structure of graphene is usually obtained in the nearest-neighbor
tight-binding model \cite{Wallace}. Conduction (or $\pi$) electrons
are allowed to jump from the $2p_z$ orbital of a carbon atom to one
of its three nearest neighbors with hopping amplitude (a.k.a. as
resonance integral) $\gamma\equiv t \approx 3$ eV. The following
dispersion relation results (see figure \ref{elecstructure}):
%%%%%%%%%%%%%%%
\begin{figure}[h]
\centering
%\includegraphics[width=10cm]{Tightbindingresltforthegraphene}
%\psfrag{3t}{$3\gamma$}
\includegraphics[width=8.5cm]{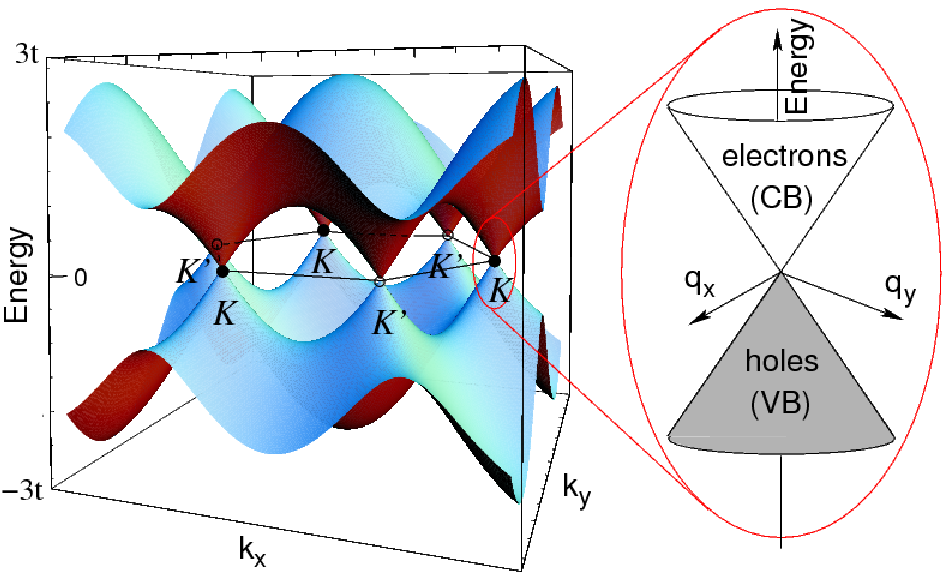}
\caption{Band structure of graphene as computed in the
tight-binding model of Wallace \cite{Wallace}. The energy $E$ is shown as a function
of the two dimensional Bloch wavevector $(k_x,k_y)$. In the vicinity
of the Dirac points at the two inequivalent corners $K$ and $K'$ of
the hexagonal Brillouin zone, the dispersion relation is linear and
hence locally equivalent to a Dirac cone (see the zoomed region). In
undoped graphene, the Fermi energy lies exactly at the two Dirac
points and the Fermi surface consists of just two points: the
valence band (VB) is filled with electrons and the conduction band
(CB) is empty. From a band theory perspective, undoped graphene is
therefore a zero gap semiconductor. However, as it is experimentally
found that it still conducts electricity at the lowest attainable
temperatures (with a conductivity of the order of a few times the
conductance quantum $e^2/h$), it is better called a (zero band
overlap) semi-metal. [Figure courtesy of Mark Goerbig]. }
\label{elecstructure}
\end{figure}
%%%%%%%%%%%%%%%%
\beq\label{eq001} E=\pm \gamma
\sqrt{\left(1+4\cos^2{(k_ya)}+4\cos{(k_ya)} \cdot
\cos{(k_x\sqrt{3}a)}\right)} \eeq where the lattice constant
$a\approx 2{.}46 $ \AA \, and the carbon-carbon distance is
$a/\sqrt{3}\approx 1.42$ \AA. Conduction (CB, $\alpha=+1$) and
valence bands (VB, $\alpha=-1$), respectively, correspond to the
different signs in the above dispersion relation ($\alpha=\pm 1$ is
called the band index); they touch at two inequivalent points --
Dirac points or valleys $K$ and $K'$ -- which are at the corners of
the hexagonal Brillouin zone in reciprocal space (see Figure
\ref{elecstructure}). These $K$ and $K'$ points are separated by a
distance $\sim 1/a$ in reciprocal space.

In the vicinity of the Dirac points the energy depends linearly on
the wave number, similarly to a massless relativistic (or
ultra-relativistic) particle (see figure \ref{elecstructure}). As a
consequence, at low energies, and close to the $K$ and $K'$ points,
the electrons can be described by a 2D massless Dirac eigenvalue equation:
%(similar to the 2D version of the Weyl equation for
%neutrinos \cite{Weyl}):
\beq\label{eq002} -i{\hbar}v_F\hat{\vec\sigma} \cdot\vec\nabla
\psi(\mathbf{r})\,=\,E\psi(\mathbf{r}) \eeq A detailed derivation of this linearized
equation starting from the tight-binding model can be found, for example, in \cite{BenaMontambaux}. Here $v_F\equiv
\sqrt{3}\gamma a/(2\hbar)\approx 10^6\ \mathrm{m/s}$ is the Fermi
velocity in graphene which plays the role of an effective velocity
of light; $\hat{\vec{\sigma}}\equiv (\hat{\sigma}_x,\hat{\sigma}_y)$
is the 2D vector of Pauli matrices,
$\psi(\mathbf{r})=(\psi_A(\mathbf{r}),\psi_B(\mathbf{r}))$ is the
two-component (bi-spinor) wavefunction of the electron, $E$ its
energy and $\vec{p}\to -i\hbar \vec{\nabla}$ is the momentum
operator in the position representation\footnote{Note that in the
low energy effective description, the wavevector $\vec{k}$ and the
corresponding momentum $\vec{p}=\hbar\vec{k}$ are now defined from
the $K$ or $K'$ point and no more from the center $\Gamma$ of the
Brillouin zone. The restriction to low energy also means that
$|\vec{k}|\ll 1/a$.}. The two components of the wavefunction refer
to the two atoms $A$ and $B$ in the unit cell. There are actually
two such Dirac equations: one for each Dirac point or valley ($K$ or
$K'$). In the following, we only consider a single Dirac cone, as if
we could separate the $K$ and $K'$ valleys. This is a valid
approximation if intervalley scattering is unlikely, which is the
case if the potential changes are smooth on the lattice scale. This
point is discussed in more detail below, see section \ref{psabig}.
The effective description in terms of massless Dirac electrons is
valid only for energies smaller than the bandwidth $\gamma \sim
3$~eV.

For a general reference on graphene see \cite{castro} and for a
pedagogical introduction see \cite{GeimKim,FuchsGoerbig}.

\subsection{Eigenstates of the 2D massless Dirac Hamiltonian}
In matrix notation, the two-dimensional massless Dirac hamiltonian
is given by: \beq\label{eq003}
\hat{H}_\textrm{kin}=\hbar{v_F}\vec{k}.\hat{\vec{\sigma}}=\hbar{v_F}
\left(\begin{array}{cccc}{0}&{k_x-ik_y}&\\{k_x+ik_y}&{0}&\end{array}\right)
\eeq where hats $\, \hat{}\, $ denote $2\times 2$ matrices in
sublattice space $(A,B)$ and the index ``kin''  refers to the
kinetic energy. For simplicity, in the following, we use units such
that $\hbar\equiv 1$ and $v_F\equiv 1$, therefore energies and
wavevectors are equal. As it will be useful when considering
piecewise constant potentials, we add a constant potential to this
hamiltonian (which simply amounts to shifting the zero of energy):
\beq\label{eq004}
\hat{H}=\hat{H}_\textrm{kin}+{V_0}\hat{1}=\left(\begin{array}{cccc}{V_0}&{k_x-ik_y}&\\{k_x+ik_y}&{V_0}&\end{array}\right)
\eeq $\hat{H}$ corresponds to the total energy,
$\hat{H}_\textrm{kin}$ to the kinetic energy and $V_0\hat{1}$ to the
potential energy, where $\hat{1}$ is the unit $2\times 2$ matrix.
From equation (\ref{eq004}) we have the eigenvalue equation:
$\hat{H}_\textrm{kin}|\psi\rangle=E_\textrm{kin} | \psi\rangle$
where $E_\textrm{kin}=E-V_0$. An eigenstate $|\psi\rangle$
corresponds to a plane wave, which we write: \beq\label{eq005}
\psi\left(\vec{r}\right)=e^{i\vec{k}.\vec{r}}\displaystyle
\binom{u}{v}=\langle \vec{r}|\psi \rangle \eeq where
$\vec{k}=(k_x,k_y)$ is the wavevector and the bispinor is given by
$u$ (respectively $v$), which is the complex amplitude on the $A$
(resp. $B$) sublattice. The corresponding kinetic energy is such
that $E_\textrm{kin}^2=k_x^2+k_y^2$. In the following, we assume
$k_y$ to be real and positive (i.e. a plane wave propagative from
left $y<0$ to right $y>0$). However, $k_x$ can be real or purely
imaginary, because $k_x^2$ can be positive or negative. If $k_x$ is
real, the wave is oscillating. If $k_x$ is purely imaginary, the
wave is evanescent.

\subsubsection{Oscillating wave}
If ${k_x}^2>0$ (i.e. $E_\textrm{kin}^2>k_y^2$), the wave is oscillating and the
wavefunction reads: %\beq\label{eq008} \psi =
%e^{i\vec{k}.\vec{r}}\frac{1}{\sqrt{2\mathcal{A}}}\displaystyle
%\binom{1}{\alpha i\sqrt{\frac{k_y-ik_x}{k_y+ik_x}}} \eeq
 \beq\label{eq009}
\psi=e^{i\vec{k}.\vec{r}}\frac{1}{\sqrt{2\mathcal{A}}}\displaystyle
\binom{1}{\alpha e^{i\phi}}=\langle \vec{r}|\vec{k},\alpha\rangle
\eeq
where $\alpha= \textrm{sgn}(E_\textrm{kin})$ is the band index ($\alpha=1$ in the
conduction band and $-1$ in the valence band) and  $\phi$ is the angle between the wave vector $\vec{k}$ and the
$x$-axis such that $\tan{\phi}=k_y/k_x$. The kinetic energy
is $E_\textrm{kin}=E-V_0=\alpha\sqrt{{k_y}^2+{k_x}^2}$,with
${k_y}^2+{k_x}^2>0$. The total surface $\mathcal{A}$ of the graphene
sheet is taken to be one in the following $\mathcal{A}\equiv 1$.
% We
%shall often use a shorter notation: \beq\label{eq009}
%\psi=e^{i\vec{k}.\vec{r}}\frac{1}{\sqrt{2}}\displaystyle
%\binom{1}{\alpha e^{i\phi}}=\langle \vec{r}|\vec{k},\alpha\rangle
%\eeq where
%In the following, most of the time, we will not need
%to bother normalizing the wavefunctions, as this has no consequences
%on such quantities as the transmission probability. In such cases,
%we will not keep track of factors such as $1/\sqrt{2\mathcal{A}}$.
%However, from time to time, the normalization will reappear when
%needed (for example, when calculating the average current or
%velocity).

\subsubsection{Evanescent wave}
An interesting possibility is that $k_x^2$ be negative (i.e.
$E_\textrm{kin}^2<k_y^2$), then $k_x=\pm i\kappa$, with
$\kappa\in\mathbb{R}^+$. We need to consider two possibilities
depending on $E_\textrm{kin}$.
\begin{itemize}
\item If $E_\textrm{kin}{\neq}0$, there are two sub-cases depending on the sign of $|k_x|$. On the one hand, if $k_x=i\kappa$
\beq\label{eq006} \psi\sim e^{ik_y{y}}e^{-\kappa x}\displaystyle
\binom{1}{\alpha i\sqrt{\frac{k_y+\kappa}{k_y-\kappa}}} \eeq and
$E_\textrm{kin}=E-V_0=\alpha \sqrt{k_y^2-\kappa^2}$, with $k_y^2-\kappa^2>0$
and $\alpha=\textrm{sgn}(E-V_0)$. This wave decays towards
increasing $x$. On the other hand, if $k_x=-i\kappa$
\beq\label{eq007} \psi\sim e^{ik_y{y}}e^{\kappa x}\displaystyle
\binom{1}{\alpha i\sqrt{\frac{k_y-\kappa}{k_y+\kappa}}} \eeq And
$E_\textrm{kin}=E-V_0=\alpha\sqrt{{k_y}^2-\kappa^2}$,with ${k_y}^2-\kappa^2>0$
and $\alpha=\textrm{sgn}(E-V_0)$.
This wave decays towards decreasing $x$.\\
\item If $E_\textrm{kin}=0$ (i.e. $k_x^2+k_y^2=0$), again two sub-cases occur. On the one hand, if $k_x=ik_y$:
\beq\label{eq010} \psi\sim e^{ik_y{y}}e^{- k_y x}\displaystyle
\binom{0}{1} \eeq On the other hand, if $k_x=-ik_y$:
\beq\label{eq011} \psi \sim e^{ik_y{y}}e^{k_y x}\displaystyle
\binom{1}{0} \eeq
\end{itemize}

These solutions will be useful in the following when considering
piecewise constant potentials.

\subsection{Potential steps and barriers in graphene}\label{psabig}
%%%%%%%%%%%%%%%
\begin{figure}[h]
\centering \psfrag{gamma}{$\gamma$}
\psfrag{V_0}{$V_0$}\psfrag{E_F}{$E_F$}\psfrag{x}{$x$}
\includegraphics[width=9cm]{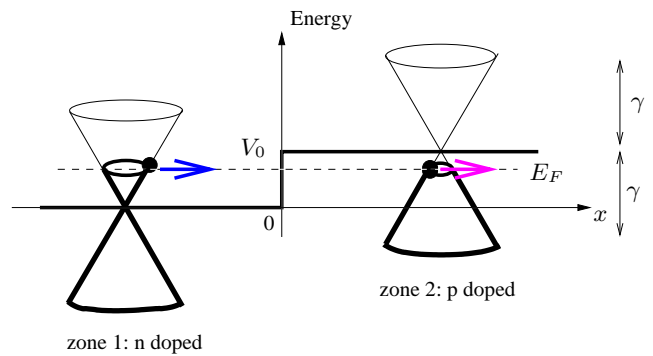}
\caption{Band structure across a square potential step $V_0$ (or
sharp $np$ junction). At equilibrium the chemical potential is
uniform $\mu(T=0)=E_F$: the Fermi level is shown as a dashed line.
The black dot represents the electron before and after the step: its
direction of motion is indicated by a blue arrow. Note that its
wavevector is reversed but not its velocity. The typical bandwidth
$\gamma\sim 3$~eV is also shown.} \label{squarestep}
\end{figure}
%%%%%%%%%%%%%%%%
We consider a potential step $V(\vec{r})$ and call $w$ the
characteristic length scale over which it varies. In graphene, it is
possible to realize potential steps that are smooth ($w\gg a$) on
the lattice scale $a\sim 0.2$ nm and therefore do not induce inter-valley
scattering as the distance between valleys in reciprocal space is
$|\vec{K}-\vec{K}'| \sim 1/a$ and the Fourier transform of the
potential $\tilde{V}(\vec{q})$ is non-zero only for $q\ll 1/a$. For
such potentials, valleys are decoupled and electrons in graphene can
be described by a single valley 2D massless Dirac hamiltonian
$\hat{H}=\hat{H}_\textrm{kin}+\hat{V}(x,y)$ \cite{ShonAndo1998}. In
addition, as the potential varies slowly over the distance between
two neighboring atoms ($A$ and $B$), the potential matrix
$\hat{V}(x,y)$ can be taken to be diagonal in the sublattice space
${V(x,y)}\hat{1}$ \cite{ShonAndo1998}. In the following, we will
consider potentials that are translationally invariant along $y$ and
therefore of the form $V(x)\hat{1}$. Therefore the complete hamiltonian reads:
\beq
\hat{H}=\hat{H}_\textrm{kin}+V(x)\hat{1}=\vec{k}\cdot \hat{\vec{\sigma}}+V(x)\hat{1}
\eeq

In addition, the potential can be smooth ($w\gg 1/k_F$) or sharp
($w\ll 1/k_F$) on the Fermi wavelength scale $1/k_F=
1/|E_\textrm{kin}|$. If it is sharp, we can assume that it is
piecewise constant (``square''). For example, $V(x)=V_0\Theta(x)$
for a square step (see figure \ref{squarestep}) and
$V(x)=V_0\Theta(x)\Theta(d-x)$ for a square barrier (see figure
\ref{barrierangle}), where $\Theta$ is the Heaviside step function.
The case of a sharp step is discussed in section \ref{sharpstep},
that of a smooth step (see figure \ref{smoothstep}) in section
\ref{smoothpotstep} and that of a sharp barrier in section
\ref{barrier}.

Depending on the doping (the position of the Fermi level), the steps and barriers
can correspond to different types of junctions. If the doping is such that the Fermi level lies
in different bands before and after the step, the junction is said to be bipolar ($np$ or $pn$
junction). If it lies in the same band, the junction is said to be unipolar ($nn'$ or $pp'$ junction). Similarly barriers can correspond to bipolar ($npn$ or $pnp$) or unipolar ($nn'n$ or $pp'p$) junctions.
In the following we will consider $np$ or $nn'$ junctions (steps) and $npn$ or $nn'n$ junctions (barriers).

\subsection{Velocity and probability current}
For later purpose, we define the average velocity and the average
probability current of an eigenstate $|\vec{k},\alpha\rangle$. Using
the Heisenberg equation of motion, the velocity operator can be
evaluated as: \beq\label{eq012} \hat{\vec{v}}\equiv
\dot{\vec{r}}=\frac{1}{i}[\vec{r},\hat{H}]=\hat{\vec{\sigma}} \eeq
From which, one can obtain the average velocity of a plane wave of
momentum $\vec{k}$ and band index $\alpha$: \beq\label{eq013}
\vec{v}\equiv \langle \vec{k},\alpha|\hat{\vec{v}}|\vec{k},\alpha
\rangle=\alpha\frac{\vec{k}}{k} \eeq Hence in the case where the
band index is $\alpha=-1$ -- i.e. when the electron is in the
valence band and has a negative kinetic energy
$E_\textrm{kin}=E-V_0<0$ -- the wavector is opposite to the
propagation of the wave (i.e. to the velocity), which is quite
unusual.

The average current is obtained in the following way. The
probability density of the state $|\psi\rangle$ is
$|\psi(\vec{r},t)|^2$ and the associated (average) probability
current is called $\vec{j}(\vec{r},t)$. The conservation of
probability \beq\label{eq020} \vec{\nabla} \cdot \vec{j} = -{\frac{
\partial}{ \partial t}} |\psi|^2 \eeq allows us to define the average current as: \beq \label{abriko}
\vec{j}=\psi^{\dagger}\hat{\vec{\sigma}}\psi \eeq
The average velocity is given by:
\beq
\langle \psi |\hat{\vec{v}} | \psi \rangle = \int d^2 r \, \psi^{\dagger}
\hat{\vec{\sigma}}\psi =\int d^2r \, \vec{j}
\eeq
For an eigenstate
$| \vec{k},\alpha\rangle$, the average current reads: \beq \vec{j}=\alpha
\frac{\vec{k}}{k} \eeq
which is just equal to the average velocity as we took a unit system area $\mathcal{A}\equiv 1$.

\subsection{Electron versus hole in the valence band}
\begin{figure}[h]
\centering
\psfrag{E_F}{$E_F$}\psfrag{E_F=0}{$E_F=0$}\psfrag{k}{$k$}\psfrag{0}{$0$}
\includegraphics[width=7cm]{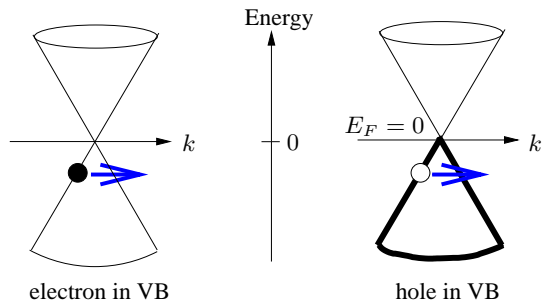}
\caption{Difference between an electron and a hole both in the
valence band (VB). We compare an electron of momentum $\vec{k}_e$ in
the VB  (on the left) with a hole that corresponds to removing an
electron of momentum $\vec{k}_e$ in an otherwise filled VB (on the
right). The electron has a negative kinetic energy $E_e=-k_e$, a
negative charge $-e$, an average velocity $\langle \vec{v}_e\rangle
= - \vec{k}_e/k_e$ opposite to its momentum and carries an electric
current $-e\langle \vec{v}_e\rangle$. The hole has a momentum
opposite to that of the electron $\vec{k}_t=-\vec{k}_e$, its kinetic
energy is positive $E_t=k_t=-E_e$, its charge is positive $+e$, its
average velocity is the same as that of the electron $\langle
\vec{v}_t\rangle = \vec{k}_t/k_t=\langle \vec{v}_e\rangle$
(therefore it moves in the same direction as the electron) and its
electric current is opposite $+e \langle \vec{v}_t\rangle$. The blue
arrow indicates the direction of motion.} \label{electronVShole}
\end{figure}

An electron moving in the valence band (VB) should not be confused
with a hole, which is the absence of an electron in an otherwise
filled band (see figure \ref{electronVShole}). Let us compare an
electron of momentum $\vec{k}_e$ in an otherwise empty VB with a
hole that corresponds to removing of an electron of momentum
$\vec{k}_e$ in an otherwise filled VB. The electron has a negative
kinetic energy $E_e=-k_e$ (as measured from the Dirac point), a
negative charge $-e$, an average velocity $\langle \vec{v}_e\rangle
= - \vec{k}_e/k_e$ opposite to its momentum and carries an electric
current $-e\langle \vec{v}_e\rangle$. Concerning the hole, its
momentum is opposite to that of the electron $\vec{k}_t=-\vec{k}_e$
\footnote{Despite the fact that on the figure \ref{electronVShole}
the hole appears to be at the same position (in reciprocal space) as
the electron, its momentum is opposite $\vec{k}_t=-\vec{k}_e$. This
is due to the fact that the hole corresponds to the \emph{removal}
of an electron. The same remark applies to the kinetic energy
$E_t=-E_e$ as measured from the Dirac point (band crossing).}, its
kinetic energy is positive $E_t=k_t=-E_e$, its charge is positive
$+e$, its average velocity is the same as that of the electron
$\langle \vec{v}_t\rangle = \vec{k}_t/k_t=\langle \vec{v}_e\rangle$
(therefore it moves in the same direction as the electron) and its
electric current is opposite $+e \langle \vec{v}_t\rangle$.

Let us now discuss Klein tunneling for an electron (initially in the
conduction band (CB)) incident on a step (see figure
\ref{squarestep}). The electron is transmitted inside the step as an
electron in the VB and not as a hole (as often stated). Conservation
of the electric current should be enough to understand that point.
Note that in order to keep moving in the same direction, the
electron has to reverse its momentum when going from the CB (outside
the step) to the VB (inside the step).
%%%%%%%%%%%%%%%
%%%%%%%%%%%%%%%%

\subsection{General conservation laws: energy $E$, momentum projection $k_y$ and 1D current $j_x$}
There are three fundamental conservation laws that we will keep on
using when considering a massless Dirac particle incident on a
potential step or barrier that is translationally invariant in the $y$ direction
$\hat{V}(x,y)=V(x)\hat{1}$. First, there is the conservation of energy $E$ as a result of time
translational invariance. Then, the momentum projection $k_y$ is also conserved
as a result of translational invariance along $y$ (i.e. parallel to the interface(s)).
Finally, since the system is time independent, the
probability conservation law reads $\vec{\nabla} \cdot \vec{j}=0$ and
translational invariance along $y$ further implies that
$\vec{j}(x,y)=\vec{j}(x)$ and therefore that the 1D current is conserved: \beq
j_x(x) =\textrm{constant} \label{eq021} \eeq

\subsection{Conservation of pseudo-spin and the absence of backscattering}\label{copataob}
\subsubsection{Pseudo-spin and chirality (or helicity)}
An electron described by the 2D massless Dirac equation carries a
pseudo-spin $1/2$ related to its freedom of belonging either to the
$A$ or to the $B$ sublattice. In graphene, the electron has still an
extra spin-type degree of freedom. It is related to its freedom of
being either close to the $K$ point or to the $K'$ point in the
Brillouin zone: this is called valley pseudo-spin. In the present
paper, we do not discuss this degree of freedom and rather
concentrate on the sublattice pseudo-spin $\vec{\sigma}$. The fact
that the electron has a sublattice pseudo-spin is encoded in its
wavefunction being a bispinor and the hamiltonian
$\hat{H}_\textrm{kin}=\vec{k}\cdot \hat{\vec{\sigma}}$ being a $2\times 2$
matrix in sublattice space. Chirality is here defined as follows. The
chirality (or helicity) operator is the projection of the sublattice pseudo-spin
operator on the momentum direction: \beq \hat{C}\equiv
\frac{\vec{k}\cdot \hat{\vec{\sigma}}}{k} \eeq Its eigenvalues are
$C = \pm 1$. When there is no potential $\hat{V}(\vec{r})=0$, the
chirality operator commutes with the hamiltonian and is therefore a
conserved quantity. The hamiltonian and the chirality can be
diagonalized by the same eigenvectors: \beq
\hat{C}|\vec{k},\alpha\rangle=\alpha |\vec{k},\alpha\rangle \eeq
which shows that the chirality $C$ is just the band index $\alpha$
in that case \footnote{In the case where both valleys $K$ and $K'$
are considered, one finds that for an eigenstate
$|\vec{k},\alpha,\xi \rangle$ of the hamiltonian the chirality
$C=\alpha \times \xi$ is the product of the band index $\alpha$ and
the valley index $\xi=+1$ (if $K$) and $-1$ (if $K'$).}.

\subsubsection{Chirality factor and the absence of backscattering} \label{abs}
Here, we discuss an important consequence of the
pseudo-spin, first discovered by Ando and coworkers in the context of
carbon nanotubes \cite{Ando1998}. Consider a massless Dirac
electron, which is incident on an impurity whose potential is smooth
on the lattice scale such that intervalley scattering is suppressed
%\footnote{This is related to the fact that if the potential is
%smooth on the lattice scale $a$, its Fourier transform is non-zero
%only for wavevectors $q\ll 1/a$ and it can therefore not provided
%enough momentum to the electron to be scattered from one valley to
%the other, because the latter are separated by a distance
%$|\vec{K}-\vec{K'}|\sim 1/a$ in reciprocal space.}
and the problem can be described within a single valley model (see section \ref{psabig}). The impurity
potential is therefore $\hat{V}_\textrm{imp}(\vec{r})\approx
U(\vec{r})\hat{1}$ \cite{Ando1998}. For simplicity, though the
argument can be made much more general (see \cite{Ando1998} and the next paragraph), we will
compute the scattering probability using the first order Born
approximation. It is given by \beq P(\theta) \propto |\langle
\vec{k'},\alpha'|U(\vec{r})\hat{1}|\vec{k},\alpha\rangle|^2 \eeq
where $|\vec{k},\alpha\rangle$ and $|\vec{k'},\alpha'\rangle$ are
the initial and the final states respectively and $\theta$ is the
angle between the final and initial wavevectors. As the collision is
elastic $k'=k$ and $\alpha'=\alpha$. Therefore the only freedom in the final state is the
angle $\theta\equiv \phi_{\vec{k'}} - \phi_{\vec{k}}$ that
$\vec{k'}$ makes with $\vec{k}$. We are now in a position to compute
the matrix element: \beq \langle
\vec{k'},\alpha'|U(\vec{r})\hat{1}|\vec{k},\alpha\rangle=\frac{1+e^{i\theta}}{2}\tilde{U}(\vec{k'}-\vec{k})
\eeq where $\tilde{U}(\vec{q})\equiv \int d^2r
U(\vec{r})\exp(i\vec{q}\cdot \vec{r})$ is the Fourier transform of
the potential $U(\vec{r})$. Note that the transferred momentum is
$q=2k\sin (\theta/2)$. Therefore, the scattering probability reads:
\beq P(\theta)\propto |\tilde{U}(\vec{q})|^2 \times
\cos^2\frac{\theta}{2} = |\tilde{U}(\vec{q})|^2 \times \frac{1+\cos
\theta}{2}\eeq The first term $|\tilde{U}(\vec{q})|^2$ is the usual
result of the Born approximation and the second
$\cos^2\frac{\theta}{2}$ is due to the sublattice pseudo-spin and is
called the ``chirality factor''. The latter is just the square of the scalar product
between the incoming and outgoing bispinors:
$(1, e^{i\phi_{\vec{k}}})/\sqrt{2}$ and $(1, e^{i\phi_{\vec{k'}}})/\sqrt{2}$.
The effect of the chirality factor is quite
dramatic as it kills backscattering ($\vec{k'}=-\vec{k}$): \beq P(\theta=\pi)\propto
|\tilde{U}(\vec{q})|^2 \times \cos^2\frac{\pi}{2} =0 \textrm{ with }
q=2k \eeq

An intuitive explanation of this absence of backscattering is the
following: if the electron tries to backscatter $\vec{k'}=-\vec{k}$
it also has to reverse its sublattice pseudo-spin $\vec{\sigma} \to
-\vec{\sigma}$ as the pseudo-spin direction is tied to that of the
momentum (indeed remember that away from the impurity
$\hat{H}_\textrm{kin}=\vec{k}\cdot \hat{\vec{\sigma}}$). However, the potential
$U(\vec{r})\hat{1}$ does not act in sublattice space (it is the unit
matrix) and can therefore not reverse the pseudo-spin. Therefore
backscattering is impossible. This has profound physical
consequences on the transport properties of massless Dirac
electrons, such as weak antilocalization
\cite{McCann2006,Bardarson2007,Nomura2007}.

\subsubsection{Conservation of pseudo-spin and the absence of backscattering}\label{cpsproof}
We now prove that a 2D massless Dirac electron normally incident on
a potential $V(x)\hat{1}$ can not be backscattered as a consequence
of the conservation of its pseudo-spin $\hat{\sigma}_x$. We assume
that the electron is initially ($t=0$) in a momentum eigenstate
$(k_x>0,k_y=0)$ and incident on a potential that is translationally
invariant in the $y$ direction ($V(x,y)=V(x)$). It is described by
the following hamiltonian: \beq \hat{H}=k_x\hat{\sigma}_x + k_y
\hat{\sigma}_y +V(x)\hat{1} \eeq The velocity operator in the $x$
direction is $\hat{v}_x=-i[x,\hat{H}]=\hat{\sigma}_x$. Its time
evolution is given by the Heisenberg equation of motion
\footnote{That the velocity operator does not commute with the
hamiltonian is peculiar to the Dirac equation and is responsible for
the so-called zitterbewegung.}: \beq
\dot{\hat{v}}_x=-i[\hat{\sigma}_x,\hat{H}]=2\hat{\sigma}_z k_y \eeq
Here, because of the translational invariance along the $y$
direction, the momentum $k_y$ is a conserved quantity:
$\dot{k}_y=-i[k_y,\hat{H}]=0$. Therefore the momentum operator along
$y$ does not evolve $k_y(t)=k_y(0)$. If the initial state of the
electron $|\psi(0)\rangle$ is an eigenstate of zero momentum in the
$y$ direction $k_y(0)|\psi(0)\rangle=0$, then at any time $t>0$:
\begin{eqnarray}
\langle\psi(t)|\dot{\hat{v}}_x(0)|\psi(t)\rangle &=&
\langle\psi(0)|\dot{\hat{v}}_x(t)|\psi(0)\rangle\nonumber \\
&=&2\langle\psi(0)|\hat{\sigma}_z(t)k_y(0)|\psi(0)\rangle=0
\end{eqnarray}
which means that the velocity (or the pseudo-spin) along $x$ is a
constant of the motion: $$\langle\psi(t)|\hat{v}_x|\psi(t)\rangle=
\langle\psi(0)|\hat{v}_x|\psi(0)\rangle=+1$$. The electron is
therefore perfectly transmitted and its motion is exactly the same
as in the absence of the potential (it is not even delayed). This
shows that a (single valley) massless Dirac electron normally
incident on a translationally  invariant potential can not be
backscattered.

An alternative explanation for the absence of backscattering (invoking supersymmetry) is presented in \cite{plyushchay}.

\section{Potential Step}\label{secpotstep}
\begin{figure}[h]
\centering
\includegraphics[width=9cm]{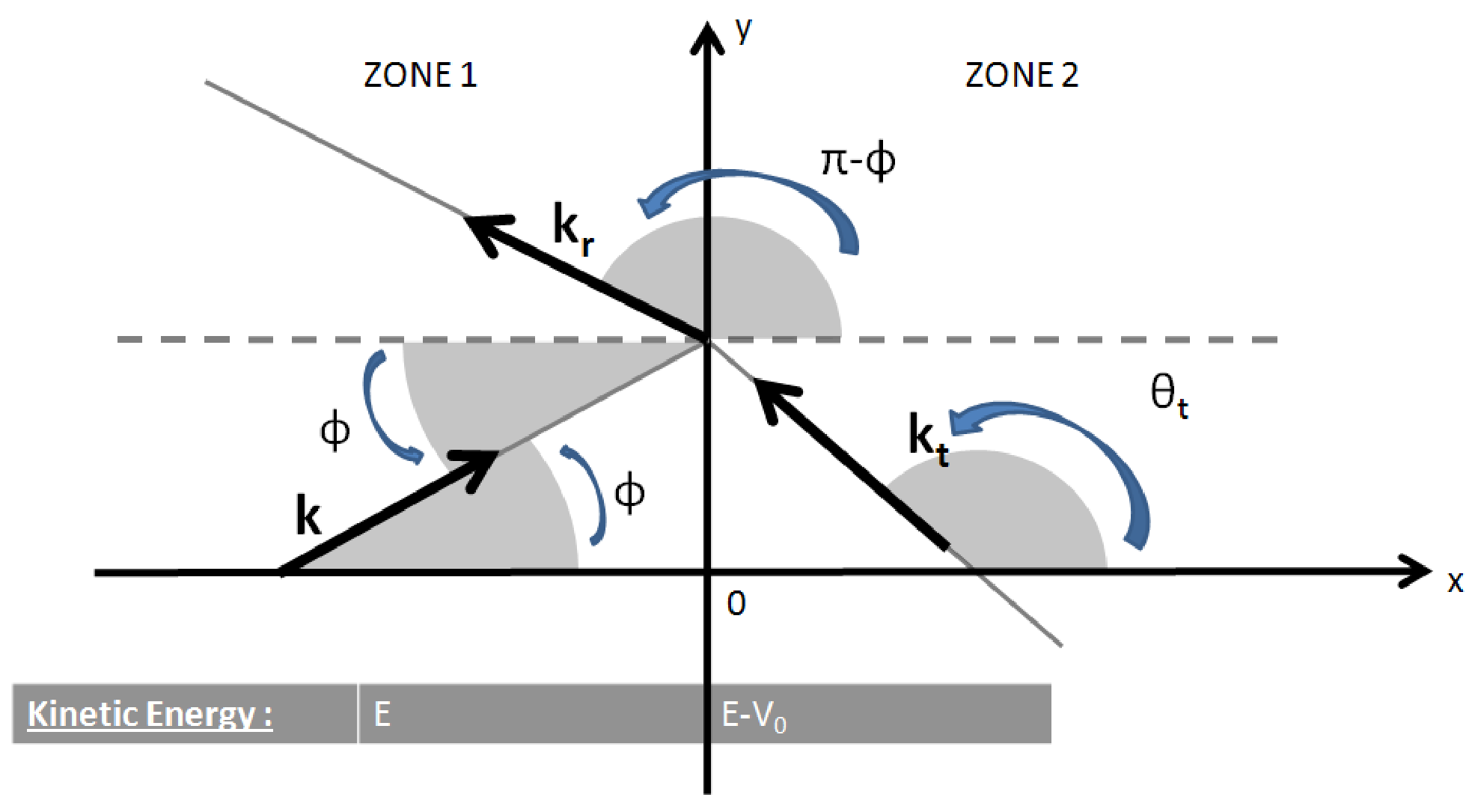}
\caption{An electron of energy $E$ is incident from the left on a
square potential step of height $V_0$ such that $V_0>E$. Angles
($\phi$ and $\theta_t$) and wavevectors in the two zones (before and
after the step) are defined as follows: incident
$\vec{k}=(k_x,k_y)=E(\cos \phi, \sin \phi)$; reflected
$\vec{k_r}=(-k_x,k_y)=E(\cos (\pi-\phi),\sin(\pi-\phi))$; and
transmitted
$\vec{k_t}=(k_x',k_y)=-(E-V_0)(\cos{\theta_t},\sin{\theta_t})$. $\phi$ is called
the incidence angle.
%The
%drawing is made for the particular case $E=V_0/2$ for which
%$\theta_t=\pi-\phi$ which is not true when $E\neq V_0/2$.
}
\label{stepangle}
\end{figure}

Let us first consider a square (sharp) potential step of height
$V_0$ on which an electron of energy $E=k_F>0$ is incident (see figure
\ref{squarestep}).
%In graphene, a $np$ junction can be realized with a
%backgate and could correspond to such a potential step for the
%electrons if it is steep enough, see \cite{huard}.
Two zones can be defined, one for $x<0$ corresponding to a kinetic energy of
$E_\textrm{kin}=E$, another for $x>0$ corresponding to a kinetic
energy of $E_\textrm{kin}=E-V_0$. An analogy with an optical system
will be made: the system is equivalent to a light beam going through
a discontinuity between two transparent media. For instance, going
from glass to air. Since $E>0$ in zone 1, $\alpha=+1$; whereas, in
zone 2, $\alpha=\textrm{sgn}{(E-V_0)}=\pm1$. We consider an incoming
electron with a given wavevector $\vec{k}$ in zone 1 and call
$\vec{k'}$ the wavevector in zone 2. Wavevectors and angles are
defined in figure \ref{stepangle}.
%%%%%%%%%%%%%%%%%%
%%%%%%%%%%%%%%%%%%%%%%
In the case where the band index is $-1$ (i.e. when $E-V_0<0$) we
saw that the momentum is opposite to the propagation of the wave
(i.e. to the velocity), because the electron moves in the valence
band. Therefore ${k'_x}<0$, see figure \ref{stepangle}. This gives
rise to anomalous refraction.

\subsection{Energy below a sharp step ($0<E<V_0$)}\label{sharpstep}
This corresponds to a $np$ junction.
From the conservation of the total energy and of the momentum
projection $k_y$, relations between angles can be deduced. The
different wavectors  are written as
$\vec{k}^{(j)}=(k_x^{(j)},k_y^{(j)})=|\vec{k}^{(j)}|(\cos
\theta_j,\sin \theta_j)$, where $E_{\textrm{kin},j}=\alpha_j
|\vec{k}^{(j)}|$ with
$j= $ \{\underline{i}ncident, \underline{r}eflected, \underline{t}ransmitted\}.
Then the conserved wavevector projection along $y$ reads
$k_y^{(j)}=\alpha_j E_{\textrm{kin},j} \sin{\theta_j}$ and we obtain
the table below:
\begin{center}
\begin{tabular}{c|c|c|c}
$j$ & incident &
reflected & transmitted \\
\hline\hline
Band index $\alpha_j$ & $1$ & $1$ &$-1$ \\
Kinetic energy $E_{\textrm{kin},j}$ & $E>0$ & $E>0$ & $E-V_0<0$ \\
Angle $\theta_j$ & $\theta_i\equiv \phi$ & $\theta_r$ & $\theta_t$
\\ \hline
\end{tabular}
\end{center}
The equality $k_y^{(i)}=k_y^{(r)}$ gives
\beq\label{eq014}\theta_r=\pi-\phi \eeq and $k_y^{(i)}=k_y^{(t)}$
gives \beq\label{eq015} E\sin{\phi}=-(E-V_0)\sin{\theta_t} \eeq

This last equation can be seen as an analog of the Snell-Descartes
equation for light refraction into a medium of negative refraction
index $n$ \cite{Falko}. To see that the refraction index is indeed
negative, one needs to define angles from the perpendicular to the
interface -- as is usual in optics. For the transmitted wave
(``refracted ray'') in zone 2, we define $\theta_2\equiv
\theta_t-\pi$ and $\theta_1\equiv \phi$ for the incident wave in
zone 1. In that case both angles $\theta_1$ and $\theta_2$ have a
modulus smaller than $\pi/2$. Hence the Snell-Descartes law now
reads ${n_1}\sin{\theta_1}={n_2}\sin{\theta_2}$ with
${n_1}\propto{E}$ and ${n_2}\propto{(E-V_0)}=E_\textrm{kin}$ where $n_2<0$.
Therefore the refraction index $n \propto E_\textrm{kin}=\alpha k$ is
proportional to the kinetic energy and may be defined as $n\equiv
\alpha ka \sim E_\textrm{kin}/\gamma$, where $\gamma\approx 3$~eV is the
hopping amplitude (or bandwidth) and $a$ is the lattice spacing.
Note that the wavevector $k\equiv |\vec{k}|$ changes from zone 1 to
zone 2, because the velocity $v_F\approx 10^6$ m/s is a constant and
the kinetic energy changes from positive to negative. Another way to
see that the refraction index is negative is to realize that the
refraction is anomalous, or in other words that the refracted ray is
closer in direction to the reflected ray than to the incident one
(see Figure (\ref{stepangle})), which is quite unusual.

In optics, the refraction index $n=c/v\propto 1/v$ is inversely
proportional to the phase velocity $v$ -- where $c\approx 3\times
10^8$ m/s is the light velocity in vacuum -- which changes when
going from one medium to another. As the photon (kinetic) energy is
unchanged upon crossing the interface, the optical index also reads
$n=k/k_0 \propto k$ where $k_0$ is the wavector in vacuum and $k$
that in the medium. In optics, media with negative index of
refraction -- so called meta-materials -- have received a lot of
attention recently, see for example \cite{Pendry}.

The next step is to write the wavefunctions in both zones 1 and 2
and to connect them on the interface at $x=0$. Here normalization of
the wavefunctions is not needed. Using equations derived in section
\ref{2dmdeig}, the wave functions can be written in both zones as
(see Figure \ref{stepangle}): \beq\label{eq016}
\psi_1=e^{ik_yy}\left[e^{i{k_x}x}\displaystyle\binom{1}{+e^{i\phi}}+re^{-i{k_x}x}\displaystyle
\binom{1}{+e^{i{(\pi-\phi)}}}\right] \eeq \beq\label{eq017}
\psi_2=te^{ik_yy}e^{i{k'_x}x}\displaystyle\binom{1}{-e^{i\theta_t}}
\eeq The continuity of the wavefunction \footnote{For the
Schr\"{o}dinger equation, we would have had to use the continuity of
the wavefunction and that of its derivative as well. Here the
two-component spinor allows the same number of equations just from
the continuity of the wavefunction.} is used in $x=0$ to obtain:
\begin{equation}
1 = -r + t \textrm{ and }e^{i\phi} =  re^{-i\phi}-te^{i\theta_t}
\end{equation}
which gives: \beq\label{eq018}
r=\frac{e^{i\theta_t}+e^{i\phi}}{e^{-i\phi}-e^{{i\theta_t}}}
\textrm{ and }
t=\frac{e^{i\phi}+e^{-i\phi}}{e^{-i\phi}-e^{{i\theta_t}}} \eeq

\begin{figure}[h]
\centering
\includegraphics[width=6cm]{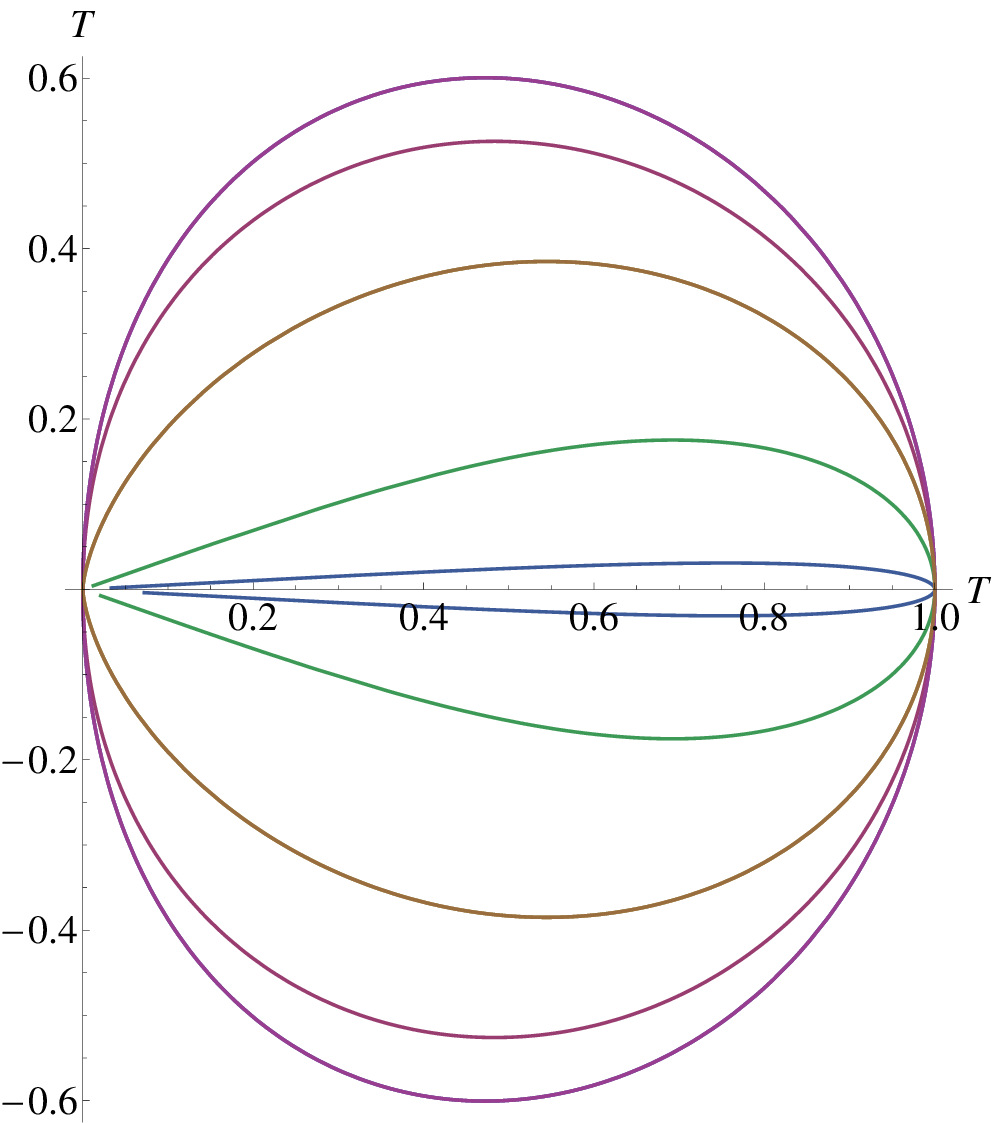}
\caption{
%Top: plot of the transmission probability $T$ as a function
%of the incident angle $\phi$ going from $-\pi/2$ to $\pi/2$ and the
%dimensionless energy $\varepsilon \equiv E/V_0$ going from $0$ to
%$1$. A critical angle exists when $\varepsilon >1/2$, where $T=0$.
%Bottom:
Polar plot of the transmission probability $T(\phi)$ for the
potential step (values of $T$ are shown on all axis) for several
dimensionless energies $\varepsilon\equiv E/V_0$ between $0$ and $1$
and with the incident angle $\phi$ running from -$\pi/2$ to $\pi/2$:
$\varepsilon=0$ (blue), $\varepsilon=0.25$ (purple),
$\varepsilon=0.5$ (beige), $\varepsilon=0.75$ (green) and
$\varepsilon=0.95$ (black). Note that for $\varepsilon=0.75$ there
is a critical angle above which $T$ is strictly equal to zero. }
\label{3dstep}
\end{figure}

We now wish to obtain the transmission probability $T$ from the
amplitudes $r$ and $t$. This requires using the conservation of the 1D current, see equation (\ref{eq021}),
which here reads:
\begin{eqnarray}
\label{eq022} &&j_x[\textrm{incident}]+j_x[\textrm{reflected}] =
j_x[\textrm{transmitted}] \nonumber \\
&\leftrightarrow& \cos \phi -|r|^2 \cos \phi = - |t|^2 \cos \theta_t
\end{eqnarray} and therefore: \beq\label{eq024}
1=|r|^2-|t|^2\frac{\cos{\theta_t}}{\cos{\phi}} \eeq This is the
probability conservation law $1=R+T$. It allows one to identify the
transmission $T$ and reflection $R$ probabilities \footnote{The
reflection coefficient $R$ is always equal to $|r|^2$, whereas the
transmission coefficient $T$ is not necessarily given by $|t|^2$. In
the present case of the potential step
$T=-|t|^2\frac{\cos{\theta_t}}{\cos{\phi}}\neq{|t|^2}$. There is a
mistake precisely on that point in \cite{Falko}. As an illustration
of the fact that $T$ is generally not given by $|t|^2$, we consider
an extreme example where the transmission probability vanishes $T=0$
although $t\neq 0$. If in zone 2, the kinetic energy vanishes
$E_\textrm{kin}=E-V_0=0$, there is an evanescent wave when $\phi \neq 0$. Then
using Eq. (\ref{eq010}) and (\ref{abriko}), we can show that
$r=\exp(i2\phi)$ and therefore $T=1-|r|^2=0$, while
$t=1+\exp(i2\phi)\neq 0$.} as: \beq\label{eq025}
T=-\frac{\cos{\theta_t}}{\cos{\phi}}{|t|^2} \textrm{ and  } R=|r|^2
\eeq and we finally obtain: \beq\label{eq026}
T=-\frac{\cos{\phi}\cos{\theta_t}}{\sin^2{\left(\frac{\phi+\theta_t}{2}\right)}}
\textrm{ and }
R=\frac{\cos^2{\left(\frac{\phi-\theta_t}{2}\right)}}{\sin^2{\left(\frac{\phi+\theta_t}{2}\right)}}
\eeq where the transmitted angle is: \beq\label{eq027}
\theta_t=\theta_2+\pi=\arcsin{\Big(\frac{E}{V_0-E}\sin{\phi}
\Big)}+\pi \eeq These equations are equivalent to the Fresnel
formulae in optics. Note that $\cos \theta_t \leq 0$ so that $T\geq
0$ as it should. The transmission probability is plotted in Figure
\ref{3dstep}.
%%%%%%%%%%%%%%%%%%%%%%%%%%
%%%%%%%%%%%%%%%%%%

The transmission coefficient vanishes beyond a certain critical
angle $\phi_c$, which is is defined as $\sin{\phi_c}=(V_0-E)/E$ (see figures \ref{3dstep} and \ref{phasediag}).
Beyond this angle, an evanescent wave is created in zone 2 and a
total reflection is observed. A critical angle only occurs when
$V_0/2<E$, otherwise the conditions to have an evanescent wave
$\frac{E}{V_0-E}\sin{\phi}>1$ and $\frac{E}{V_0-E}\sin{\phi}<-1$ are
never satisfied. This is similar to the optical phenomenon of total
internal reflection. For instance if a light beam goes from a
glass-like medium with an index $n_1$ to an air-like medium of index
$n_2<n_1$ then using the Snell-Descartes law
${n_1}\sin{\theta_1}={n_2}\sin{\theta_2}$, there exists a refracted
beam only if the incident angle $\theta_1$ is smaller than a
critical angle $\theta_c=\arcsin{({n_2}/{n_1})}$.

If the electron arrives at normal incidence ($\phi=0$) the
probability to go through is 1: \beq\label{eq031}
T\left(\phi=0\right)=1 \eeq This is due to the ``absence of
backscattering'' discussed previously ( see section \ref{abs} and
Ref. \cite{Ando1998})  and is a consequence of the pseudo-spin
conservation of the massless Dirac electron. This (at first)
surprising result is most often referred to as the ``Klein tunnel
effect'' although it is not a genuine tunnel effect in the quantum
mechanical sense. Indeed it involves no classically forbidden region
and no evanescent wave. It is a consequence of (1) the existence of
negative kinetic energy states (valence band) in the step, that
match the positive kinetic energy states (conduction band) outside
the step and (2) of the conservation of pseudo-spin which permits
the transition. This second point is quite important. The chirality
factor acts as a selection rule. Indeed the square of the scalar
product between the incoming and the transmitted bispinors is
$[1+\alpha \alpha' \cos (\theta_t-\phi)]/2=1$ as
$\alpha=+1=-\alpha'$, $\theta_t=\pi$ when $\phi=0$. However, it is
not because states are available in the step at the matching energy
that the transition will necessarily happen. The case of the bilayer
graphene (with its massive chiral electrons) is illuminating in this
respect \cite{Katsnelson}. In a graphene bilayer, the carriers are
also described by a bispinorial wavefunction, but with a Berry phase
of $2\pi$ (instead of $\pi$ for the monolayer). The low energy
effective hamiltonian is \beq
\hat{H}_\textrm{bilayer}=-\frac{\hbar^2}{2m^*}[(k_x^2-k_y^2)\hat{\sigma}_x+2k_xk_y
\hat{\sigma}_y] \label{bilayer}\eeq for a single valley, where $m^*$
is the effective mass of the carriers. As the band structure is also
that of a gapless semiconductor -- although with parabolic bands
$$E=\pm (\hbar \vec{k})^2/(2m^*)$$ -- there are also states of
negative kinetic energy available in the step. Here, however, the
pseudo-spin conservation forbids the interband transition at normal
incidence $T(\phi=0)=0$. Indeed the chirality factor \footnote{Note
the factor of two difference in the cosine. This is a consequence of
the Berry phase being $2\pi$ rather than $\pi$.} in this case is
$[1+\alpha \alpha' \cos 2(\theta_t-\phi)]/2=0$ as
$\alpha=+1=-\alpha'$, $\theta_t=\pi$ when $\phi=0$. The 1D case is
studied in an appendix.

Some special set of energy $E$ and incidence angle $\phi$ are worth
mentioning. For instance, if $E=V_0/2$ one has (see also
\cite{Cheianov}): \beq\label{eq030} T=\cos^2{\phi}=1-(k_y/k_F)^2
\eeq and if $E\ll V_0$ (because $\theta_t\to \pi$): \beq
T=\frac{2\cos \phi}{1+\cos\phi} \eeq These two cases are quite
interesting (see figure \ref{3dstep}). There is no critical angle,
and the transmission is always quite large unless $\phi$ becomes
really close to $\pm \pi/2$. There is a slight preference for normal
incidence but no true collimation effect. This will later be
compared to the case of a smooth step.

\subsection{Energy above a sharp step ($V_0<E$)}
This corresponds to a $nn'$ junction.
For $x<0$ the kinetic energy is $E$ and for $x>0$ the kinetic energy
is $E-V_0>0$. Here we will consider both $V_0>0$ and $V_0<0$. The
wavefunctions are: \beq\label{eq032} \psi_1=e^{ik_yy}\left [
e^{i{k_x}x}\displaystyle\binom{1}{+e^{i\phi}}+e^{-i{k_x}x}\displaystyle
\binom{1}{+e^{i{(\pi-\phi)}}}\right ] \eeq \beq\label{eq033}
\psi_2=te^{ik_yy}e^{i{k'_x}x}\displaystyle\binom{1}{+e^{i\theta_t}}
\eeq and the Snell-Descartes relation reads: \beq\label{eq034}
E\sin{\phi}=(E-V_0)\sin{\theta_t} \eeq This leads to a new system of
equations that can be solved to give: \beq\label{eq035}
r=\frac{e^{i\phi}-e^{i\theta_t}}{e^{-i\phi}+e^{i\theta_t}} \eeq
\beq\label{eq036}
t=\frac{e^{-i\phi}+e^{i\phi}}{e^{-i\phi}+e^{i\theta_t}} \eeq and
one eventually obtains \beq\label{eq037}
T=|t|^2\frac{\cos{\theta_t}}{\cos{\phi}}=\frac{\cos{\phi}\cos{\theta_t}}{\cos^2
{\left(\frac{\phi+\theta_t}{2}\right)}} \eeq where:
\beq\label{eq038}
\theta_t=\arcsin{\Big(\frac{E}{E-V_0}\sin{\phi}\Big)} \eeq When
$V_0>0$, this case corresponds to refraction indices $n_1>0$ and
$n_2>0$ (equivalent to usual optical materials: the refraction is
normal)  with $n_1>n_2$, so that there is also a critical angle
given by $\sin \phi_c=(E-V_0)/E$. This is typically similar to a
light beam going from glass to air.  In classical mechanics, there
is a similar phenomenon. Consider a non-relativistic particle of
mass $m$ and energy $E=\vec{p}^2/(2m)$ incident on a potential step
of energy $V_0$ such that $V_0<E$. The conservation of energy
$p^2/(2m)={p'}^2/(2m)+V_0$ and that of parallel momentum $p'_y=p_y$
imply that ${p'}_x^2/(2m)=p_x^2/(2m)-V_0$. The particle will
therefore be reflected if ${p'}_x^2<0$ because a negative kinetic energy is
classically forbidden. This defines a critical angle of incidence
$\phi_c$ given by $\sin{\phi_c} = \sqrt{(E-V_0)/E}$, where the angle
$\phi$ is defined by $\tan{\phi}=p_y/p_x$. We note that the
non-relativistic refraction index $n\propto \sqrt{E_\textrm{kin}}\propto
|\vec{p}|$ differs from the ultra-relativistic one of graphene
$n\propto E_\textrm{kin}=\alpha |\vec{k}|$.

When $V_0<0$, this case corresponds to refraction indices $n_1>0$
and $n_2>0$ with $n_1<n_2$, so that there is no critical angle. This
is typically similar to a light beam going from air to glass.

Using the Schr\"odinger equation (parabolic band approximation), the fact that a
particle with an energy above a potential step ($E>0>V_0$) and with
an incident velocity in the same direction as the step force
($F=-\nabla V$) can be reflected is known as a ``quantum
reflection'', because it can not be understood from a classical
perspective. It is particularly spectacular for perpendicular
incidence ($\phi=0$) and low energy $E\ll|V_0|$ where the
transmission probability vanishes $T\to 0$. In the present case of a
massless Dirac electron, quantum reflection is absent as
$T(\phi=0)=1$ for any energy $E\geq 0$. This is again a consequence
of the absence of backscattering.

\subsection{Smooth potential step}\label{smoothpotstep}
%%%%%%%%%%%%%%%
\begin{figure}[h]
\centering
\psfrag{V_0/2}{$V_0/2$}\psfrag{E_F}{$E_F=0$}\psfrag{x}{$x$}\psfrag{w}{$w$}
\psfrag{-w}{$-w$}\psfrag{-V_0/2}{$-V_0/2$}\psfrag{V_0}{$V_0$}\psfrag{2w}{$2w$}
\includegraphics[width=8cm]{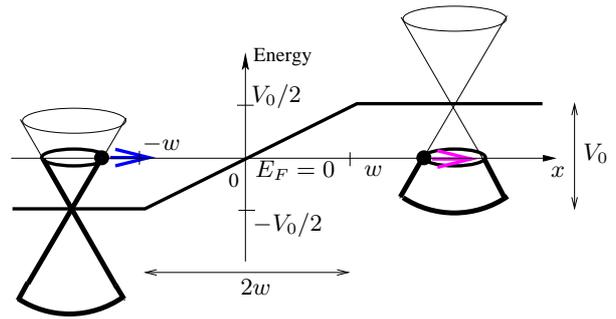}
\caption{Band structure across a smooth $np$ junction. The potential
height is $V_0$ and its width is $2w$. The energy of the incoming
electron (Fermi energy) is half that of the potential step. The zero
of energy is chosen to coincide with the Fermi level. Arrows
indicate the direction of motion before (blue) and after (pink) the step.} \label{smoothstep}
\end{figure}
\begin{figure}[h]
\centering
\includegraphics[width=7cm]{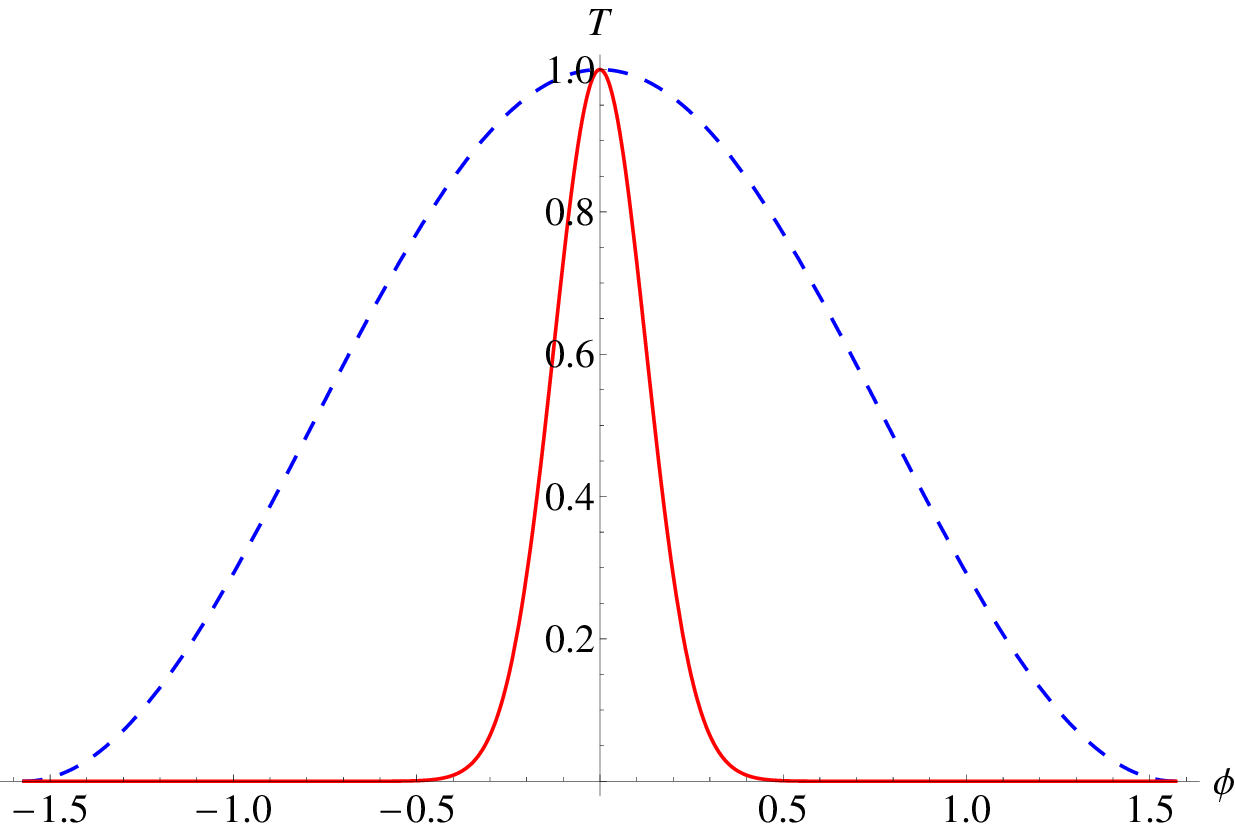}
\caption{Transmission probability $T(\phi)$ across a potentiel step
for an energy which half that of the barrier. $T$ (on the $y$ axis)
is plotted as a function of the incidence angle $\phi$ (on the $x$
axis) between $-\pi/2$ and $\pi/2$ for a sharp (dashed blue,
$T=\cos^2\phi$) and a smooth step (red, $T=e^{-\pi k_F w
\sin^2\phi}$ with $k_Fw=10$). The collimation effect is clearly seen
in the smooth step case, with a characteristic angle
$\phi_0=1/\sqrt{\pi k_Fw}\approx 0.2$.} \label{smoothsharp}
\end{figure}

%%%%%%%%%%%%%%%%
We now turn to the case of a smooth potential step, which was
recently considered by Cheianov and Falko in the context of a $np$
junction in graphene \cite{Cheianov}. Take a smooth potential step
of height $V_0$, which occurs over a distance $2w$ (Figure
\ref{smoothstep}). Smoothness roughly means $\lambda_F \ll w$
(below, we discuss a refined criterion). The potential is taken to
be $V(x)=\textrm{sign}(x) V_0/2$ when $|x|>w$ and $V(x)=Fx$ when
$|x|<w$, with $F=V_0/(2w)$. We consider a symetric situation: that
of an incoming massless Dirac electron with an energy which is half
that of the step. This energy is here $E=E_F=0$ as a result of the
shift in the zero of energy -- indeed $V(x<-w)=-V_0/2$. The electron
has a wavevector $(k_x,k_y)=k_F(\cos \phi, \sin \phi)$ where
$k_F=V_0/2$. A normally incident electron $k_y=0$ is perfectly
transmitted as a consequence of the absence of backscattering. We
focus on the case $k_y\neq 0$ which gives rise to a classically
forbidden zone close to $V(x)=E$ (i.e. $x=0$). It can be found from
the conservation of energy $\sqrt{k_x(x)^2+k_y^2}+Fx=E=0$ and the
requirement that $k_x(x)^2<0$. This defines the region
$|x|<l_c\equiv k_y/F$. The size of the classically forbidden zone is
$2l_c=2w \sin \phi$. An electron with $k_y\neq 0$ that wishes to go
through the step has to tunnel via evanescent waves to the other
side across the classically forbidden zone. In the semiclassical
approximation, the probability amplitude to tunnel is given by
$A\approx e^{iS}$ where $S=\int_{-l_c}^{+l_c} dx\, k_x(x)$ is the
(reduced) action, which is imaginary when
$k_x(x)=i|k_x(x)|=i\sqrt{k_y^2-(Fx)^2}$. The tunneling probability
is then $T=|A|^2$. The typical $k_x(x)$ is $k_x(0)=ik_y$, therefore
$S\sim i k_y\times 2l_c$ and $T \sim e^{-4 k_y l_c}$. More
precisely, $S= ik_y l_c \int_{-1}^{1} du \sqrt{1-u^2}=i\pi k_y
l_c/2$ and the probability is \cite{Cheianov}: \beq T(\phi)\approx
e^{-\pi k_y l_c}=e^{-\pi k_y^2/F}=e^{- \pi k_F w \sin^2 \phi} \eeq
This result is valid for a smooth step and for incidence angles
$\phi$ not too close to $\pi/2$. It satisfies $T(0)=1$ and
$T(\phi)\approx 1$ for incidence angles such that $|\phi|\ll
\phi_0\equiv \sqrt{F/\pi k_F^2}=1/\sqrt{\pi k_F w}$ -- where $\phi_0$ is called the collimation angle --, and then
rapidly goes to zero for oblique incidence $|\phi|> \phi_0$. This
shows that a single $np$ junction has a collimation effect as it
focuses the electronic flow by allowing the transmission of only the
trajectories that are close to normal incidence \cite{Falko}. This
result should be compared to $T=\cos^2 \phi$ found previously in the
case of a sharp step (see figure \ref{smoothsharp}).
%%%%%%%%%%%%%%%
%%%%%%%%%%%%%%%%
Apart from small incidence angles, chirality seems not to play an
important role here. This tunneling is similar to interband
tunneling in a Zener diode, see e.g. \cite{aronov}. In a
semiconductor, one can have a junction between an electron-doped and
a hole-doped region (a so-called $np$ junction). If the voltage
difference across the tunnel junction is strong enough, it is
possible for electrons to tunnel from the conduction to the valence
band across the junction. The only difference here is that the
semiconductor is 2D, gapless and the bands are linearly dispersing.

The transition between the abrupt ($w\to 0$) and smooth ($w \to
\infty$) steps limit is treated in \cite{cayssol}. Actually, a step
is considered smooth when the semiclassical approximation is valid.
This happens when the typical tunneling action is large $|S|\sim
k_x(0)l_c = k_yw \sin \phi=k_F w \sin^2 \phi \gg 1$.
The smoothness criterion is therefore $k_Fw\sin^2 \phi \gg 1$.
It not only depends on $k_Fw$ but also on
the incidence angle $\phi$. Close to normal incidence, any step
becomes sharp (see figure 1b in Ref. \cite{cayssol}).  If $k_F w<1$, the step is sharp at any incidence angle.
If $k_Fw>1$, the step is sharp close to normal incidence and smooth close to grazing incidence. In the ``sharp
step angular region'' ($|\phi|\ll 1/\sqrt{\pi k_F w}$), the transmission
is high and corresponds to the absence of backscattering; whereas in the ``smooth step
angular region'' ($|\phi|> 1/\sqrt{\pi k_F w}$), transmission is exponentially suppressed
and occurs via evanescent wave inter-band tunneling. The signature of chirality for the smooth step is therefore in the collimation effect: only electrons sufficiently close to normal incidence are transmitted across the junction.

When the energy of the incoming electron is not half that of the barrier,
the transmission probability becomes \cite{huard}:
\beq
T(\phi)\approx \exp(- \pi \frac{2k_{F1}^2}{k_{F2}+k_{F1}} w
\sin^2 \phi)
\eeq
where $k_{F1}$ (resp. $k_{F2}$) is the Fermi wavevector in the zone 1 (resp. zone 2). The symetric case considered above corresponds to $k_{F1}=k_{F2}=k_F$.

\section{Potential barrier}\label{barrier}
Let us now consider a square potential barrier \cite{Katsnelson}, see also \cite{ChenTao}. We distinguish three zones: zone
1 for $x<0$, where the potential is equal to $0$; zone 2 for
$0<x<d$, where the potential is equal to $V_0$; and zone 3 for
$x>d$, where the potential is again 0. And study two cases: $0<E<V_0$
(coresponding to a $npn$ junction)
and $E>V_0>0$ ($nn'n$ junction).
%%%%%%%%%%%%%%%%%
\begin{figure}[h]
\centering \psfrag{0}{$0$}\psfrag{E_F}{$E_F$}\psfrag{x}{$x$}
\psfrag{d}{$d$}\psfrag{V_0}{$V_0$}
\includegraphics[width=8cm]{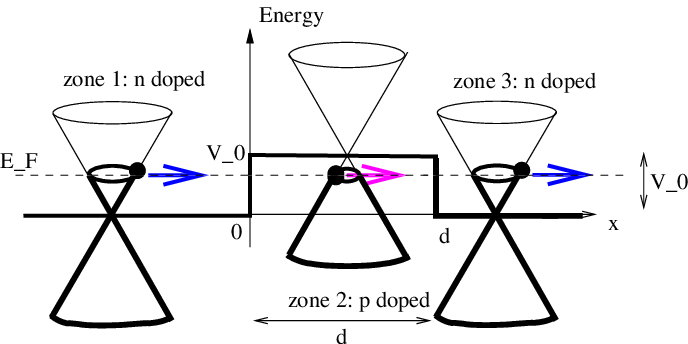}
\includegraphics[width=8cm]{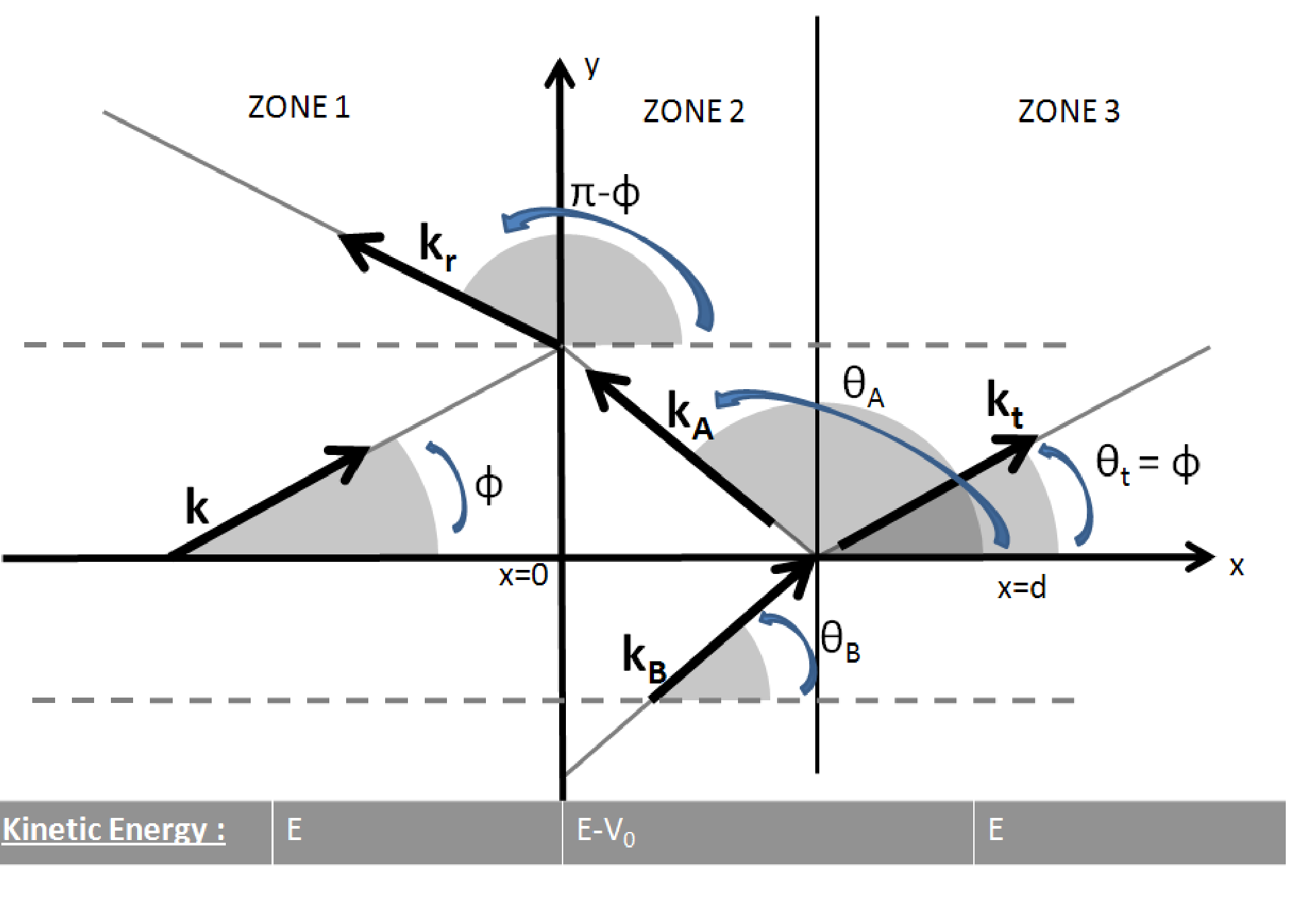}
\caption{Top: Band structure across a sharp $npn$ junction. An
electron of energy $0<E<V_0$ is incident from the left on a square
potential barrier of height $V_0$ and width $d$. Note the position
of the Dirac cone tips in the three zones.  Bottom: definition of
the angles and the wavevectors in the three zones: incident
$\vec{k}=(k_x,k_y)=E(\cos{\phi},\sin{\phi})$, reflected
$\vec{k_r}=(-k_x,k_y)=E(\cos{(\pi-\phi)}, \sin{(\pi-\phi)})$,
transmitted inside the barrier $\vec{k_A}=(k'_x,
k_y)=-(E-V_0)(\cos{\theta_A}, \sin{\theta_A})$, reflected inside the
barrier $\vec{k_B}=(-k'_x, k_y)=-(E-V_0)(\cos{\theta_B},
\sin{\theta_B})$ and transmitted $\vec{k_t}=\vec{k}$.
%The drawing is
%here made for the particular case $E=V_0/2$ for which
%$\theta_B=\phi$ and $\theta_A=\pi-\phi$ which is not true when
%$E\neq V_0/2$.
} \label{barrierangle}
\end{figure}
%%%%%%%%%%%%%

\subsection{Energy below the barrier ($0<E<V_0$)}
This corresponds to a $npn$ junction in graphene.
Using the equations of section \ref{2dmdeig} the wave functions can
be written in the three zones (see figure \ref{barrierangle}):
\beq\label{eq042}
\psi_1=e^{i{k_y}y}\left[e^{i{k_x}x}\displaystyle\binom{1}{+e^{i\phi}}+re^{-i{k_x}x}\displaystyle
\binom{1}{-e^{-i\phi}}\right] \eeq \beq\label{eq043}
\psi_2=e^{i{k_y}y}\left[Ae^{i{k'_x}x}\displaystyle\binom{1}{-e^{i\theta_A}}+Be^{-i{k'_x}x}\displaystyle
\binom{1}{e^{-i\theta_A}}\right] \eeq \beq\label{eq044}
\psi_3=te^{ik_yy}e^{i{k_x}x}\displaystyle\binom{1}{e^{i\phi}} \eeq
With the continuity of the spinors in $x=0$ and $x=d$, the following
system is obtained:
\begin{eqnarray}
1 & = & -r +  A + B \\
e^{i\phi} & = & re^{-i\phi}-Ae^{i\theta_A}+Be^{-i\theta_A} \\
0 & = & -te^{ik_xd}+Ae^{i{k'_x}d}+Be^{-i{k'_x}d} \\
0 & = & -te^{i\phi}e^{ik_xd}-Ae^{i\theta_A}e^{i{k'_x}d}+Be^{-i\theta_A}e^{-i{k'_x}d} \\
\nonumber \
\end{eqnarray}
where $A$, $B$, $r$ and $t$ are complex amplitudes to be determined.
The easiest method to solve this system is by substitution and only
$r$ is worth computing \cite{Katsnelson}:
\begin{eqnarray}\label{eq051}
r&=&-2e^{i\phi}\sin{({k'_x}d)}\times
\\
&\times&\frac{\sin{\phi}+\sin{\theta_A}}{e^{-i{k'_x}d}\cos{(\phi+\theta_A)}+e^{i{k'_x}d}\cos{(\phi-\theta_A})+2i\sin{({k'_x}d})}\nonumber
\end{eqnarray} The transmission coefficient follows from $T=1-|r|^2$:
\beq\label{eq052}
T=\frac{\cos^2{\phi}\cos^2{\theta_A}}{\cos^2{\phi}\cos^2{\theta_A}\cos^2{({k'_x}d)}+\sin^2{({k'_x}d)}[1+\sin{\theta_A}\sin{\phi}]^2}
\eeq where
${k'_x}d=-2{\pi}l\sqrt{1-2\varepsilon+\varepsilon^2\cos^2{\phi}}$
with the dimensionless
barrier width $l\equiv V_0 d/(2\pi\hbar v_F)=V_0 d/(2\pi)$  and the dimensionless
energy $\varepsilon \equiv{E/V_0}$. Naturally, from the conservation of $k_y$ we can obtained
the Snell-Descartes law: \beq\label{053}
E\sin{\phi}=-(E-V_0)\sin{\theta_A} \eeq  Several cases are worth
investigating:

\subsubsection{Low energy ($E\ll V_0$)}
The incoming electron has an tiny energy compared to $V_0$ and the
transmission coefficient takes a simpler form \cite{Katsnelson}:
\beq\label{054}
T=\frac{\cos^2{\phi}}{1-\cos^2{({k'_x}d)}\sin^2{\phi}} \eeq

\subsubsection{Grazing energy ($E\rightarrow{V_0}^{-}$)}
Here, the electron arrives exactly with the energy of the barrier.
The transmission is entirely via evanescent waves (except exactly at
$\phi=0$). When $E_\textrm{kin}=0$, $k_x'=ik_y$ and therefore $1/\cos
\theta_A=|E_\textrm{kin}|/k_x'=0$ and $\tan \theta_A=k_y/k_x'=-i$. Therefore
equation (\ref{eq052}) becomes: \beq\label{055}
T=\frac{\cos^2{\phi}}{\cosh^2{({k_y}d)}-\sin^2{\phi}} \eeq where we
used that $$\cos^2{(ik_yd)}=\cosh^2{(k_yd)} \textrm{ and }
\sin^2{(ik_yd)}=-\sinh^2{(k_yd)}$$. Note that it is not obvious a
priori that a formula obtained for oscillating waves remains valid
in a regime of evanescent waves.

\subsubsection{Normal incidence ($\phi=0$)}
If the incident angle is zero (the angle is taken from the
$x$-axis), the transmission coefficient is exactly equal to $1$
regardless of the length $d$ and the (energy) height $V_0$ of the
barrier. This perfect transmission at normal incidence is again due
to the conservation of the pseudo-spin leading to the absence of
backscattering (see section \ref{copataob}). In particular, it is not an
interference effect between the two interfaces at $x=0$ and $x=d$.
For such an effect, see section \ref{fpr} on Fabry-P\'erot
resonances.

\subsubsection{Critical angle ($1/2<\varepsilon<1$)}
\begin{figure}[h]
\centering
\includegraphics[width=7cm]{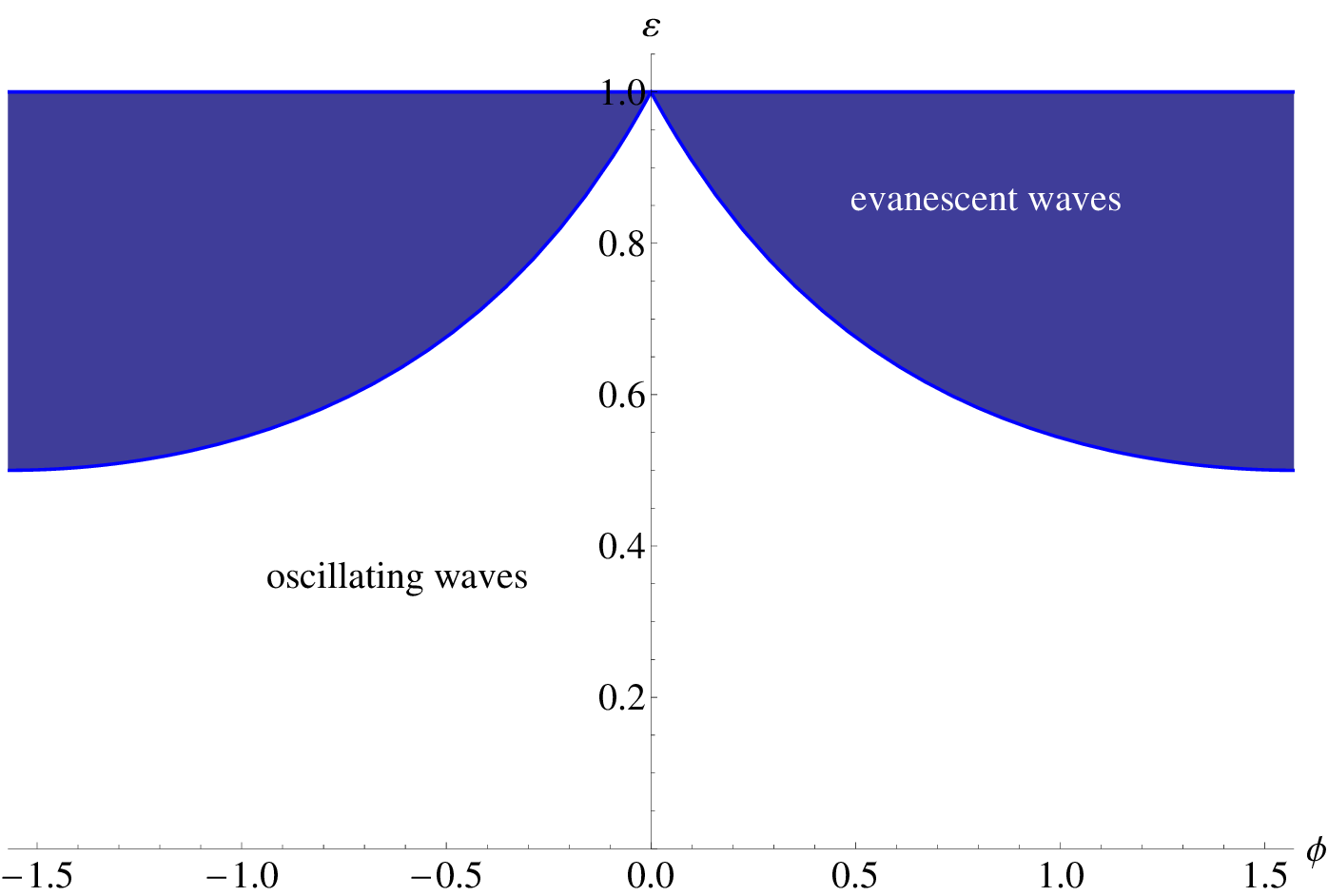}
\caption{``Phase diagram'' for the wave nature inside the barrier
plotted in the ($\phi,\varepsilon$) plane. On the $y$-axis is the
dimensionless energy $\varepsilon\equiv E/V_0$, on the $x$-axis is
the incident angle $\phi$ running from -$\pi/2$ to $\pi/2$. The
white area is the zone of oscillating waves (classically allowed
region; possibility of resonances), the blue area is the zone of
evanescent waves (classically forbidden; possibility of true tunnel
effect via an evanescent wave). The two regions are separated by the
critical angle line $\phi_c$ such that $\sin
\phi_c=(1-\varepsilon)/\varepsilon$.} \label{phasediag}
\end{figure}

As with the potential step, once a critical angle defined as
$\sin{{\phi}_c}=(V_0-E)/E$ is reached, an evanescent wave is present
in the zone 2. But contrary to what happened with the step, there is
no total reflection. Indeed, the existence of a second discontinuity
between the zone 2 and the zone 3 allows the wave to be transmitted
through the barrier with a reduced amplitude, just like in the
Schr\"{o}dinger tunnel effect: passing through is classically
forbidden but quantum mechanically permitted via an evanescent wave.
This is therefore a genuine tunnel effect. The condition to have an
evanescent wave is that the wavevector projection along $x$ in zone
2 is purely imaginary: ${k'_x}^2<0$ \footnote{Note that the
condition for having an evanescent wave in the barrier is the same
as that for total reflection on the potential step and defines the
critical angle. As $\phi$ runs between $-\pi/2$ to $\pi/2$, the
function $\sin{\phi}$ increases with $\phi$ and $\cos{\phi}\geq{0}$.
Therefore $\phi>\phi_c$ with
$\sin{\phi_c}=\frac{1-\varepsilon}{\varepsilon}{\Leftrightarrow}\sin{\phi}>\sin{\phi_c}=\frac{1-\varepsilon}{\varepsilon}$,
hence
$\varepsilon^2\sin^2{\phi}>{(1-\varepsilon)^2=1+\varepsilon^2-2\varepsilon}{\Leftrightarrow}{1-2\varepsilon+\varepsilon^2\cos{\phi}}<0$,
which is precisely the same as $k_x'^2<0$.}. As ${k'_x}^2{d^2}=(2\pi
l)^2(1-2\varepsilon+\varepsilon^2\cos{\phi})$, the condition to have
an evanescent wave is: : \beq\label{056}
1-2\varepsilon+\varepsilon^2\cos{\phi}<0 \eeq There are no
evanescent waves for an electron with an energy below $V_0/2$. The
critical angle is plotted in figure \ref{phasediag} in a kind of
``phase diagram''.
%%%%%%%%%%%%%%%%%%%
%%%%%%%%%%%%%%%%%%%%%%

\subsubsection{Fabry-P\'erot resonances ($\phi\neq 0$)}\label{fpr}
\begin{figure}[h]
\centering
\includegraphics[width=5cm]{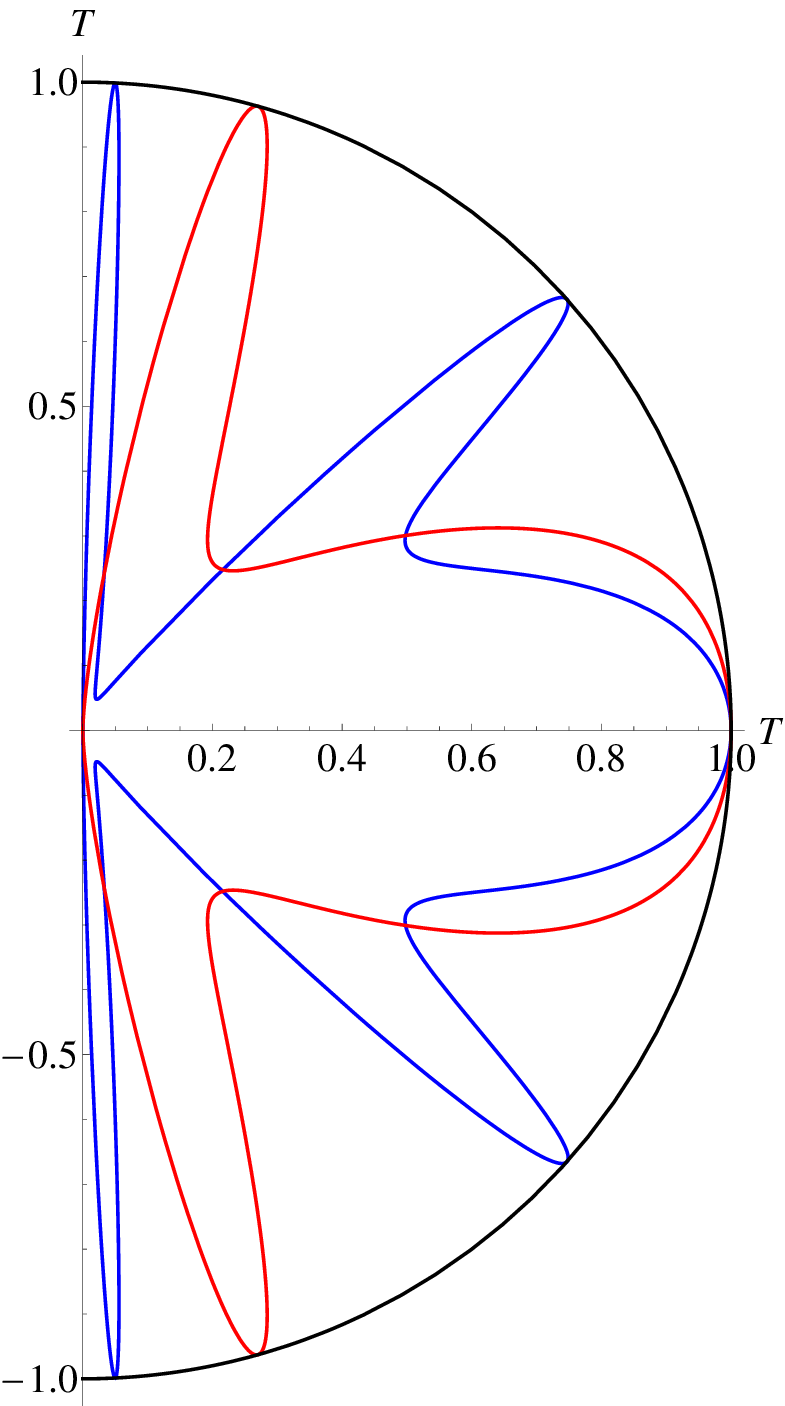}
\caption{Polar plot of the transmission coefficient $T(\phi)$ for
the potential barrier at fixed energy $\varepsilon$ and width $l$.
The two sets of parameters are the same as in \cite{Katsnelson}
namely: ($\varepsilon=0.41519$, $l= 4.85$, blue) and
($\varepsilon=0.291038$, $l=6.91$, red). The petal structure is
clearly seen and correspond to Fabry-P\'erot resonances. Note that
$T(0)=1$ as a consequence of pseudo-spin conservation and
independently of $\varepsilon$ and $l$. The black line indicates
unit transmission.} \label{petals}
\end{figure}

\begin{figure}[h]
\centering
\includegraphics[width=9cm]{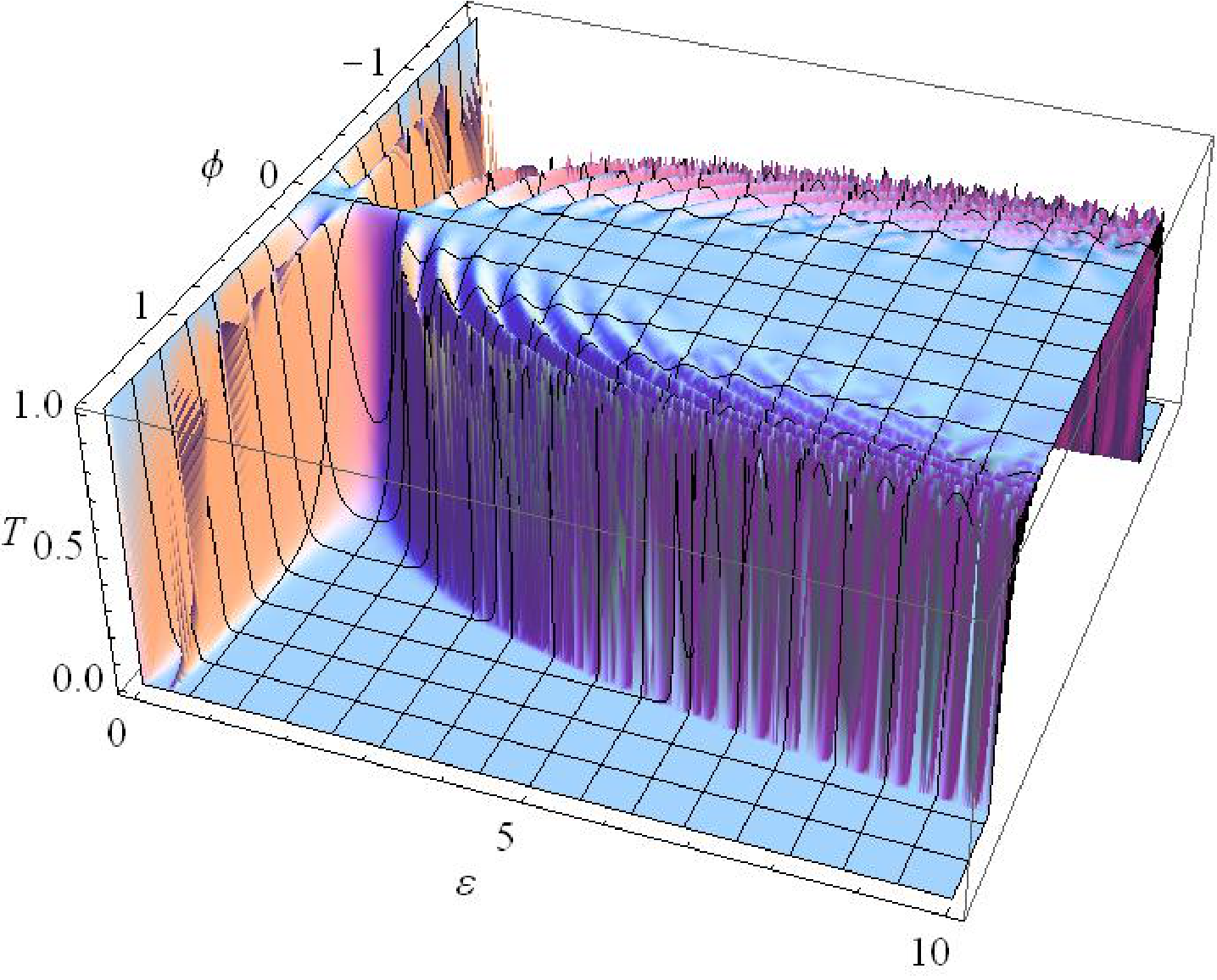}
\caption{Plot of the transmission probability $T$ as a function of
the incident angle $\phi$ and the dimensionless energy
$\varepsilon\equiv E/V_0$ for a fixed dimensionless width of the
barrier $l=1$. The energy $\varepsilon$ varies between 0 and 10 and
the angle $\phi$ between $-\pi/2$ and $+\pi/2$. Fabry-P\'erot
resonances are clearly visible.
%Those three sets are:($l=1$,$\phi\in{[-\pi/2,\pi/2]},\varepsilon\in{[0,10]}$),
%($l=4.85$,$\phi\in{[-\pi/2,\pi/2]},\varepsilon\in{[0,2]}$),($l=1$,$\phi\in{[-\pi/2,\pi/2]},\varepsilon\in{[0,2]}$).
%The first graph shows that for a greater energy, $T\rightarrow{1}$.
%On the three graphs, resonances are visible.There are more
%resonances if $l$ is greater (compare graph 2 and 3).
}
\label{3dbarrier}
\end{figure}

We now restrict to oblique incidence ($\phi\neq 0$) and consider
multiple interferences effects. A potential barrier can be seen as a
double interface (at $x=0$ and $x=d$) and as the analog of a
well-known optical system: a Fabry-P\'{e}rot interferometer. The
cavity is the region inside the barrier, which can accommodate
oscillating waves -- especially at $\varepsilon<1/2$, see Fig.
\ref{phasediag}. Accordingly, the incoming wave might interfere with
itself between the two interfaces (at $x=0$ and $x=d$) in zone 2. If
the waves interfere constructively transmission resonances will
occur where $T(\phi\neq 0)=1$. The condition of such resonances --
also known as tunneling resonances -- is \cite{Katsnelson}:
 \beq\label{057}
{k'_x}d={\pi}\times \textrm{integer}\eeq which is just the condition
that a half-integer ($\mathbb{N}/2$) number of wavelengths
($2\pi/k_x'$) along $x$ fits in the cavity (i.e. inside the barrier)
of size $d$: $(\textrm{integer/2})\times(2\pi/k_x')=d$. As
$k_x'd=-2\pi l \sqrt{1-2\varepsilon+\varepsilon^2\cos^2\phi}$, the
resonance condition involves the energy $\varepsilon$, the length of
the barrier $l$ and the angle $\phi$ and reads: \beq
2l\sqrt{1-2\varepsilon+\varepsilon^2\cos^2\phi}=\textrm{integer}
\eeq It defines specific angles $\phi_n\neq 0$ such that
$T(\phi_n)=1$. These Fabry-P\'{e}rot resonances are responsible for
the petal-like shape of $T(\phi)$ when plotted as a function of the
incidence angle $\phi$ at fixed energy $\varepsilon$ and barrier
width $l$ (see figures \ref{petals} and \ref{3dbarrier}, and Ref.
\cite{Katsnelson}).
%%%%%%%%%%%%%%%%%%%%%%%
%%%%%%%%%%%%%%%%%%%%%%
%%%%%%%%%%%%%%%%%%%%%%%%%%%
%%%%%%%%%%%%%%%%%%%%%%%%%%%%%%%

\subsection{Energy above the barrier ($E>V_0>0$)}
We now focus on the situation where an electron is incident with an
energy larger than the one of the barrier (corresponding to a $nn'n$
junction). After proceeding with the same kind of computation as for
the other case, we obtain a transmission coefficient which has the
same expression as (\ref{eq052}) and (\ref{053}) with the
replacement $\theta_A \to -\theta_A$. Within this energy range, most
of the properties of the transmission coefficient are retrieved. In
the high energy limit ($E\gg V_0>0$) limit, we find the transmission
probability: \beq\label{058}
T=\frac{\cos^2{\phi}}{1-\cos^2{{k'_x}d}\sin^2{\phi}} \eeq

\subsection{Grazing energy ($E=V_0>0$)}
Both limits $E\rightarrow{V_0}^{\pm}$ have the same common value,
therefore $T(E={V_0})=T(E={V_0}^{\pm})$: \beq \label{cosh}
T=\frac{\cos^2{\phi}}{\cosh^2{({k_y}d)}-\sin^2{\phi}} \nonumber \eeq
This result plays a key role in the calculation of the two-terminal
conductance of a ballistic undoped graphene sheet using the
Landauer formula (see \cite{kat2} and \cite{tworzydlo}). The
reservoirs are assumed to be made of doped graphene and
the sample is undoped. In order to
ensure that the reservoirs conduct much better than the undoped
graphene sheet, one takes the limit of a very large doping of the
reservoirs (many transverse channels in the reservoirs): this is
equivalent to $\sin \phi \to 0$. In this limit ($\phi \to 0$), the
transmission coefficient becomes: \beq\label{059} T \approx
\frac{1}{\cosh^2{({k_y}d)}} \eeq This corresponds to a genuine
tunnel effect through the sample (except at $\phi=0$). Indeed, when $k_yd\gg 1$,
$T\approx 4\exp(-2k_yd) \ll 1$,  which is of the expected
semiclassical form with a tunneling probability proportional to
$\exp(-2S_i)$, where $S_i=k_yd$ is the imaginary time
action for the classical path inside the barrier.

%The result (\ref{cosh}) was used in \cite{kat2,tworzydlo} to compute
%the two-terminal conductance $G_2$ of an undoped ballistic graphene sheet.

%To be continued with the calculation done in Cargese.....

%Do also the case with $V_0<0$ (potential well) and $E<0$ and $E>0$??? Bound states???

\vspace{0.2cm}

To conclude this section on the potential barrier, we mention that the case of a smooth (trapezoidal) barrier has been treated in \cite{sonin}.
See in particular the discussion of Fabry-P\'erot resonances.

\section{Experiments}\label{exps}
Below we describe the current status of transport experiments
designed at observing the Klein tunnel effect in graphene and rely
mainly on the following papers \cite{huard,gorbachev,stander,kim},
but mention also other relevant experiments
\cite{lemme,williams,barbaros,liu}. For reading convenience, in this
section, we restore the units of $\hbar$ and $v_F$.

\subsection{Backgate and topgate}
Graphene samples are usually equiped with a backgate that allows one
to electrically control the doping of the sheet through an electric
field effect (similar to a capacitor) \cite{Novoselov}. In
exfoliated samples, the backgate is usually made of doped silicon
and separated from the graphene sheet by $\sim 300$ nm of silicon
dioxide (dielectric). Electrostatic potential steps and barriers can
be made by using an additional topgate. The simultaneous use of a
backgate and a topgate allows one to control independently the
energy of incoming electrons ($E_F$) and the step/barrier height
($V_0$). The distance between the topgate and the graphene sheet
roughly gives the step size $2w$. Steps made in this way are smooth
on the lattice scale ($w\gg a$, as an example $2w\sim 80$ nm
\cite{huard}) and until recently also smooth on the Fermi wavelength
scale ($ w > 1/k_F$, typically $1/k_F \sim 10$ nm). Sharp steps ($w
< 1/k_F$) should soon become available thanks to rapid progress in
fabrication\footnote{Reaching small $k_F$ values is limited by the
presence of so-called electron-hole puddles close to the Dirac point
in a graphene sheet. These inhomogeneities mean that locally the
Fermi wavevector is never really zero but has a finite minimal value
. The cleanest suspended samples correspond to $1/k_F \sim 100$
nm.}. The typical barrier width $d$ is in between 100 nm and 1
$\mu$m.

Transport measurement are performed on these systems, in which the
two-terminal resistance is measured as a function of the backgate and topgate
voltages. Roughly speaking, the backgate controls the Fermi energy
and the topgate sets the barrier properties (height $V_0$, width $d$
and step size $2w$). One may in addition apply a magnetic field in
order to measure the magneto-resistance.

\subsection{Ballistic versus diffusive regime}
The samples are usually small: the distance $L$ between two
measuring contacts is typical less than 1 $\mu$m. This is done in
order to be as close as possible to the ballistic regime, in which
collisions on impurities can be neglected (note that our complete
discussion of Klein tunneling assumed that we could neglect the
effect of disorder). To know whether this is indeed the case implies
to compare the mean free path $l_m$ with the step size $2w$, the
barrier width $d$ and the sample size $L$. Depending on the
situation, it is possible to have a sample which is, for example,
globally diffusive ($l_m\ll L$), that the barrier is diffusive
($l_m\ll d$) but that each step can be described as being ballistic
($l_m\gg w$). The mean free path in a good graphene sample is
typically on the order of $l_m \sim 100$ nm. The effect of disorder
on a graphene $np$ junction is discussed in \cite{fogler}. These
authors find that the transition between ballistic ($\beta\gg 1$)
and diffusive ($\beta \ll 1$) regimes is controlled by a single
dimensionless parameter $\beta \equiv |dn/dx|/n_i^{3/2}$, where
$dn/dx$ is the density gradient right at the junction and $n_i\equiv
e/h \mu_m$ -- where $\mu_m$ is the mobility -- roughly gives the
density of impurities. This prediction was confirmed experimentally
\cite{stander}.

Low temperature is also needed in order to have coherent propagation
of the electrons (no decoherence). This corresponds to our
assumption of treating the electron as ideal matter waves, rather
than classical particles. The coherence length in a graphene sample
is typically $L_\phi \gtrsim 1\, \mu$m when the temperature is below
4 K.

\subsection{Smooth $np$ junctions and poor screening}
Potential steps realized up to now were generally of the smooth
type. However, the slope $F$ of the potential step $V(x)\approx Fx$
right at the junction (when $V(x)=E_F$) is not properly estimated as
$V_0/2w$ (as we did in section \ref{smoothpotstep}). Such an
estimate relies on assuming perfect screening in the graphene sheet,
which is not correct. Indeed, close to the bipolar junction,
screening in graphene is very poor because it corresponds to the
crossing of the Dirac point where the density of states vanishes.
Linear Thomas-Fermi screening would predict no screening at all, as
the inverse screening radius vanishes. It is however possible to
study non-linear screening in this regime. This was done by Zhang
and Fogler for a ballistic $np$ junction \cite{zhang}. They find
that the slope $F$ is strongly enhanced compared to the naive
estimate $V_0/2w$ (typically by a factor of 10). This effect is
important to take into account when comparing theory and experiment
\cite{stander,kim}.

\subsection{Evidences for the observation of Klein tunneling}
\subsubsection{Resistance of a smooth ballistic $np$ junction}
The first experiments designed at observing Klein tunneling in
graphene all measured the resistance across a $npn$ junction as a
function of topgate and backgate voltages
\cite{huard,gorbachev,stander}. Because of the presence of disorder,
the resistance of the barrier was found to be correctly described as
the sum of the resistance of two smooth $np$ junctions (in series).
Each $np$ junction was in the ballistic regime but not the whole
barrier. As the momentum of an electron is not conserved during its
motion between the two interfaces, cavity type resonances are not
possible, see e.g. \cite{Rossi}. We therefore consider these experiments as testing the
resistance $R$ across a smooth step ($np$ junction) rather than a
barrier.

In order to set the stage, we give typical values of relevant
quantities. The barrier height $V_0\sim 0.1$~eV, width $d\sim
300$~nm and step size $2w \sim 100$ nm; the Fermi wavelength $1/k_F
\sim 10$ nm such that $k_Fw\sim 5$ (smooth step); the mobility $\mu_m
\sim 1000-10000$~cm$^2$/V.s, the mean free path $l_m\sim 30-100$~nm
and the length between measuring contacts $L\sim 1.3-5$~~$\mu$m.

These measurements probe the average transmission across the
junction. Indeed, the two terminal conductance $G=1/R$ is given by
the Landauer formula
\begin{eqnarray}
G_{np}&=&4\frac{e^2}{h}\sum_{ch.} T_{ch.} \nonumber \\
&\approx& 4\frac{e^2}{h} \int _{-k_F}^{k_F} \frac{d k_y}{2\pi/W} T(k_y)
\end{eqnarray}
where the sum is over transverse channels labelled by $k_y=k_F\sin
\phi$ and with transmission probability $T(k_y)$, $W$ is the sample
width (not to be confused with the step size $w$) and the factor $4$
accounts for valley and spin degeneracy in graphene.

As an example, we consider a symmetric $np$ junction: the Fermi
energy is at half the potential step. On the one hand for a smooth
step of potential slope $F$ at the junction, the transmission is
$T(k_y)=e^{-\pi \hbar v_F k_y^2/F}$ (see section
\ref{smoothpotstep}) and therefore \cite{Cheianov}: \beq
G_{np}^\textrm{smooth}=\frac{2e^2}{\pi h}W\sqrt{\frac{F}{\hbar v_F}}
\eeq On the other hand, for a sharp symmetric step, the transmission
is $T=\cos^2 \phi=1-(k_y/k_F)^2$ (see section \ref{sharpstep}) and
therefore the conductance is: \beq
G_{np}^\textrm{sharp}=\frac{8e^2}{3 h}\frac{k_FW}{\pi} \eeq For
comparison, when there is no step, the transmission is perfect
$T(k_y)=1$ and the conductance is \beq G_\textrm{no
step}=4\frac{e^2}{h} \frac{k_F W}{\pi} \eeq As
$G_{np}^\textrm{sharp}=2G_\textrm{no step}/3 \gg
G_{np}^\textrm{smooth}$, we conclude that the sharp $np$ junction is
almost transparent, while the smooth $np$ junction is highly
resistive and only lets the electrons close to normal incidence
through (collimation effect). Note that from the measurement of
$G_{np}^\textrm{smooth}$ and $G_\textrm{no step}$ it is possible to
estimate the collimation angle as
$G_{np}^\textrm{smooth}/G_\textrm{no step}\sim \phi_0/2$.

The resistance measured across smooth $np$ junctions as a function
of the topgate voltage (controlling the barrier properties) was
found \cite{gorbachev,stander} in agreement with theory provided the
ballistic regime is reached ($\beta \gg 1$) \cite{fogler} and the
poor screening in graphene close to the junction is acounted for
\cite{zhang}. Quite counter-intuitively\footnote{A widespread
misconception about Klein tunneling is that it should systematically
allow electrons to go through any barrier with a high probability.
If this was true, it would indeed seem counter-intuitive that an
experimental evidence for Klein tunneling comes from measuring an
\emph{increase} of the resistance. Note that this increase is here
defined with respect to a diffusive model and not to a situation in
which Klein tunneling would be simply turned off. And that the
evidence for Klein tunneling in such an experiment on smooth
junctions is in the collimation effect.}, the resistance across a
smooth junction was measured to \emph{exceed} that predicted in a
purely diffusive model (i.e. excluding chirality effects) but to
agree with the ballistic prediction. The signature of Klein
tunneling was precisely found in this collimation effect that only
allows electrons close to normal incidence to go through the smooth
junction.

Although the average transmission was found to agree with the
prediction of Klein tunneling across a smooth $np$ junction, the
angular dependence of the transmission probability $T(\phi)$ was not
seen. In particular, from a measurement of the average transmission,
it is not possible to tell that perfect tunneling occurs at normal
incidence or to precisely measure the collimation angle.
A measurement of the resistance across a sharp ballistic
junction would be closer to revealing perfect tunneling as its
resistance is predicted to be only $3/2$ times larger than in the
absence of a step. In conclusion, these first experiments gave indirect
evidences of Klein tunneling in graphene bipolar junctions.

\subsubsection{Conductance oscillations and magneto-resistance across a ballistic $npn$ junction}
A second type of experiment was performed in order to test Klein
tunneling more directly, which relied on quantum interferences
between two $pn$ interfaces. A graphene $npn$ junction was realized
with a narrow topgate such that the whole barrier was in the
ballistic regime \cite{kim}. Indeed, the mean free path was
estimated as $l_m \gtrsim 100$ nm larger than the barrier width $d<
100$ nm. The step of size $2w\sim 30$ nm was smooth compared to both
the lattice spacing and the Fermi wavelength $1/k_F\sim 4$ nm. The
mobility was $\mu_m \sim 5000$~cm$^2$/V.s, the distance between
measuring contacts $L\sim 3$~$\mu$m and the typical barrier height
$V_0\sim 0.3$~eV. Two main observations were made on this system.

First, oscillations in the conductance as a function of the
top gate voltage revealed that the whole $npn$
junction was (for the first time) in the ballistic regime. These
oscillations where interpreted as interferences due to multiple
reflections between the two $pn$ interfaces (see the section
\ref{fpr} on Fabry-P\'erot resonances). The topgate allows one to
tune these transmission resonances and to span the interference
fringes. Note that, due to perfect tunneling at normal incidence, Fabry-P\'erot resonances are
only possible for oblique ($\phi \neq 0$) trajectories.

Second, applying a perpendicular magnetic field, resistance
measurements revealed a half-period shift in these Fabry-P\'erot
fringes above a critical magnetic field $\sim 0.3$ T. The
interpretation is as follows \cite{Levitov}. Reflectionless
transmission at normal incidence ($\phi=0$) -- in other words, Klein
tunneling -- also means that the reflection amplitude $r$ undergoes
a $\pi$ phase jump when the incident angle $\phi$ goes from positive
to negative value\footnote{In section \ref{sharpstep}, we showed
that, in the case of a sharp potential step, the reflection
amplitude is
$r(\phi)=(e^{i\phi}+e^{i\theta_t})/(e^{i\phi}-e^{-i\theta_t})$ where
$\theta_t=\pi +\arcsin (E\sin\phi/(V_0-E))$. In the limit were
$\phi=\pm \eta$ with $\eta \to 0^+$, it follows that $r(\pm
\eta)\approx e^{\pm i \pi/2} \eta V_0/(2(V_0-E))$. Therefore, there
is a $\pi$ phase jump (in the reflection amplitude) when the
incident angle changes sign: $\textrm{Arg}\, r(\eta)-\textrm{Arg}\,
r(-\eta)\approx \pi$ when $\eta \to 0^+$. More generally, for
arbitrary $\eta$, one can show that $\textrm{Arg}\,
r(\eta)-\textrm{Arg}\, r(-\eta)=\pi+2[\eta+\arcsin(E\sin \eta
/(V_0-E))]=\pi +\mathcal{O(\eta)}$. }. At zero magnetic field, two
consecutive (non-normal) reflections on the two $pn$ interfaces
occur with opposite angles $\phi_1$ and $\phi_2=-\phi_1$. A weak
magnetic field bends the electronic trajectories. Above a critical
field, trajectory bending becomes sufficient to make the two
consecutive reflections occur with the same incident angle
$\phi_1=\phi_2$. This suddenly adds $\pi$ to the phase accumulated
by an electron between two reflections and shifts the interference
fringes by half a period. The observation of this half-period shift
(see figure 3 in \cite{kim}) is therefore a direct evidence of
perfect tunneling at normal incidence.

\section{Conclusion}\label{ccl}
%In conclusion, we compare the well-known quantum
%mechanical tunnel effect (of the Schr\"{o}dinger equation) and the
%Klein tunnel effect just presented (related to the massless Dirac
%equation). For the Klein tunnel effect, there are two energy bands
%which translates in the fact that the Dirac equation has a $2\times 2$ matrix
%structure whereas the Schr\"{o}dinger equation is scalar. Moreover,
%the electronic dispersion relation is gapless (the two band touch)
%and linear $E\propto{|\vec{k}|}$. In contrast, in conventional
%semiconductors (modeled with the Schr\"{o}dinger equation) the gap
%between the valence and conduction band is usually finite,
%only the partially filled band is considered (conduction band) and the dispersion relation
%is quadratic $E\propto{|\vec{k}|}^2$. These two facts -- 1) two bands
% 2) that touch -- imply the existence of \emph{classically reachable}
%states of negative kinetic energy. This is a key point in
%understanding the Klein tunnel effect for massless electrons: an
%electron incident from the conduction band (positive kinetic energy)
%can enter a barrier by accessing a classically allowed state in the
%valence band (negative kinetic energy) without quantum mechanical
%tunneling. The Klein tunnel effect is therefore not a genuine ``tunnel
%effect'' in the usual sense of quantum mechanics: it does not
%involve passing trough a classically forbidden region via an
%evanescent wave.
In conclusion, we compare Klein tunneling to the standard tunnel
effect and outline what are
the crucial ingredients necessary for its occurence. We also
show that the phrase ``Klein tunneling'' has different meanings.
Eventually we give indications for further reading.

The standard tunnel effect across a barrier is an intraband transition
of a Schr\"odinger electron via evanescent waves across the
classically forbidden zone (the barrier). It gives a tunneling
probability at normal incidence which is roughly $T(\phi=0)\sim
e^{-2\kappa d}$ where $i\kappa$ is the typical wavevector ( in the $x$ direction, perpendicular to the barrier)
inside the barrier and $d$ is the barrier width. The probability therefore decays
exponentially with the width and the energy height of the
barrier (through $\kappa$).

Klein tunneling is the name given to the interband transition (say
from a conduction to a valence band) across a step or barrier of a
massless Dirac electron. It relies (i) on having negative kinetic energy
states available in the step or barrier (matching energy of the incoming electron)
and also (ii) on the pseudo-spin conservation that may allow or not the transition (roughly speaking this is a kind of selection rule
given by the overlap of the bispinors outside and inside the barrier: its modulus square is the so-called chirality factor).

Depending on the precise situation Klein tunneling may
refer to different physical situations and mechanisms. Here we distinguish four
situations encountered in the present article (by default, it is
usually the first case which is meant):

1) At normal incidence on a potential step that is translationally
invariant (along $y$), there is perfect interband transmission
without evanescent waves: $$T(\phi=0)=1$$ This is a consequence of
the absence of backscattering due to pseudo-spin $\hat{\sigma}_x$
conservation. It is not a genuine quantum tunnel effect. See section
\ref{cpsproof}.

2) At oblique incidence on a sharp step, there can be $T<1$
interband transmission without evanescent waves. For example, for an
electron incident with an energy which is half that of the step:
$$T(\phi\neq 0)=\cos^2 \phi $$
The chirality factor is playing a role here. See section \ref{sharpstep}.

%These two cases 1) and 2) could be called \emph{chiral tunneling}.

3) At oblique incidence on a smooth step, there is interband
tunneling (via evanescent waves). For example, for an electron
incident with an energy which is half that of the step:
$$ T(\phi\neq 0) \approx e^{-\pi \hbar v_F k_F^2 \sin^2 \phi/F}$$
where $F$ is the potential gradient at the bipolar junction. This is
a genuine quantum tunnel effect when $|\phi|>\phi_0\equiv
\sqrt{F/(\pi \hbar v_F k_F^2)}$ and chirality only plays a role
close to normal incidence leading to collimation of the electrons
($|\phi | \ll \phi_0$). See section \ref{smoothpotstep}.

4) At oblique incidence on a square barrier, there can be
Fabry-P\'erot resonances. These are transmission resonances due to
the multiple interferences of oscillating waves between the two
interfaces:
$$ T(\phi_n\neq 0)=1 \textrm{ when } k_x' d =\pi n$$
See section \ref{fpr}.

We end this article by giving a list of references for further
reading on topics not covered in the present review. Klein tunneling
has also been studied theoretically for massive Dirac electrons in
``gapped graphene'' \cite{SetareJahani}, for massive chiral
electrons in a graphene bilayer with \cite{Sim} or without
\cite{Katsnelson} band gap and in a deformed honeycomb lattice
\cite{Omri}. It was also investigated in monolayer graphene in the
presence of a magnetic field \cite{Cheianov,Levitov,Masir} or a
superlattice \cite{LouieCohen,Barbier}. The motion of a wavepacket through a
barrier in graphene is discussed in \cite{Pereira}.

\begin{acknowledgement}
We thank the participants of the Carg\`ese summer school of
mesoscopics (october 2008, GDR-CNRS Physique quantique
m\'esoscopique organised by B. Reulet, Ch. Texier and G.
Montambaux), D. Jahani, M. Goerbig, G. Montambaux, Ch. Texier and
especially F. Pi\'echon for many interesting discussions. And also
M. B\"uttiker for encouragements and S. Gu\'eron, B. Huard and P. Carmier for
useful comments on the manuscript. This work was realized during the
internship (June-July 2008) of P.E.A. at LPS Orsay in partial
fulfillment of his master (M1) degree at the Universit\'e Paris-Sud,
France.
\end{acknowledgement}

\appendix
\section{Klein tunneling in one-dimension}\label{app}
As an illustration of the importance of pseudo-spin conservation, we
consider Klein tunneling in 1D and study two toy-model hamiltonians.
First consider the following massless Dirac hamiltonian -- which we
call the ``1D monolayer'': \beq \hat{H}_m=k_x \hat{\sigma}_x
+V(x)\hat{1} \label{1dm} \eeq with $\hbar v_F\equiv 1$ where $v_F$ is the Fermi
velocity. The pseudo-spin is conserved as
$[\hat{\sigma}_x,\hat{H}_m]=0$, and thus also the velocity operator
$\hat{v}_x=-i[x,\hat{H}_m]=\hat{\sigma}_x$ is a conserved quantity
\footnote{Among Dirac equations, this is peculiar to the 1D case,
that features no zitterbewegung.}. If the electron is initially in a
velocity eigenstate (say such that $v_x=+1$), then $\langle
\hat{v}_x(t)\rangle=+1$ at any $t>0$. Therefore the motion of the
electron in the presence of the potential is exactly the same as in
its absence (the motion is not even delayed). This is a strong
consequence of the absence of backscattering.

To understand the physical meaning of the pseudo-spin, we consider the case when the potential is absent.
An eigenstate of the
hamiltonian is: \beq
\psi_{k_x,\sigma_x}(x)=\frac{1}{\sqrt{2}}\displaystyle
\binom{1}{\sigma_x}e^{i\sigma_x k_x x}
 \eeq
where $k_x$ is the momentum, $E_\textrm{kin}=\textrm{sign}(E_\textrm{kin})|k_x|$ is the
energy and $\sigma_x=\pm 1$ is the eigenvalue of the pseudo-spin
$\hat{\sigma}_x$. The latter can also be written as
$\sigma_x=\textrm{sign}(k_x)\textrm{sign}(E_\textrm{kin})$ or $\sigma_x=E_\textrm{kin}/k_x$ and is
therefore the direction of motion (+1 for right movers and -1 for
left movers). If an electron is initially a right mover, it will
remain so even if it encounters regions of arbitrary non-zero
potential.

In the presence of the potential, it is also possible to find the
eigenstates of the hamiltonian (\ref{1dm}), see \cite{Bocquet}.
Performing a unitary transformation the hamiltonian can be written
as $\tilde{H}_m=k_x\hat{\sigma}_z+V(x)\hat{1}$. The eigenvalue
equation decouples in two equations $\mp i d \psi_\pm /dx
=[E-V(x)]\psi_\pm$, which are easily solved to give the following
eigenvectors \beq \psi_{E,\sigma_x}(x)\sim
\frac{1}{\sqrt{2}}\displaystyle \binom{1}{\sigma_x}e^{i\sigma_x
\int^x dx' [E-V(x')]} \eeq at any eigenenergy $E$. Each energy level
is doubly degenerate as $\sigma_x=\pm 1$. This clearly shows that,
for a scalar potential $V(x)\hat{1}$ at any energy, the eigenstates
are delocalized as $|\psi_{E,\sigma_x}(x)|^2=$ constant.
%%%%%%%%%%%%%%%%%%%%%%%%%%%
\begin{figure}[h]
\centering
\psfrag{x}{$x$}\psfrag{Energy}{Energy}
\includegraphics[width=5cm]{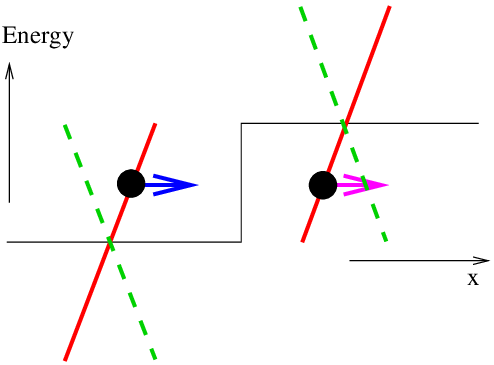}
\includegraphics[width=5cm]{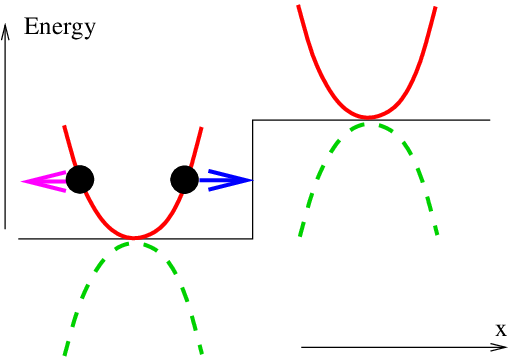}
\caption{Klein tunneling in one dimension. An electron incident from
the left on a sharp potential step (the blue arrow indicates its
direction of motion). Top: the ``1D monolayer'' case, in
which the pseudo-spin corresponds to the direction of motion: the branch of right movers is in red $\sigma_x=+1$
 and that of left movers in dashed green $\sigma_x=-1$. Bottom: the ``1D bilayer'' case, in
 which the pseudo-spin corresponds to the band index: the conduction band is in red $\sigma_x=+1$
 and the valence band in dashed green $\sigma_x=-1$. In both cases, conservation of pseudo-spin imposes the
 direction of motion after the step (indicated by a pink arrow).} \label{1dmono1dbi}
\end{figure}
%%%%%%%%%%%%%%%%%%%%%%%%%%%%%%%

Next, consider the case of massive chiral electrons (a kind of ``1D
bilayer'' toy-model) described by the following hamiltonian: \beq
\hat{H}_b=k_x^2 \hat{\sigma}_x +V(x)\hat{1} \eeq with
$\hbar^2/(2m^*)\equiv 1$ where $m^*$ is the effective mass (see
equation (\ref{bilayer}) for the corresponding 2D hamiltonian). The
pseudo-spin is again a conserved quantity. However, the velocity
operator is not, as it is now given by
$\hat{v}_x=-i[x,\hat{H}_b]=2k_x \hat{\sigma}_x$ and
$[\hat{v}_x,\hat{H}_b]=-2i\hat{\sigma}_x dV/dx\neq 0$. When $V(x)=0$, the
eigenvectors of $\hat{H}_b$ are: \beq
\psi_{k_x,\sigma_x}(x)=\frac{1}{\sqrt{2}}\displaystyle
\binom{1}{\sigma_x}e^{i\sigma_x k_x x} \eeq with the corresponding
eigenenergies $E_\textrm{kin}=\sigma_x k_x^2$, which shows that here
$\sigma_x=\textrm{sign}(E_\textrm{kin})$ can also be seen as the band index.
When $V(x)$ is non zero, we can perform a unitary transformation to
rewrite the hamiltonian as
$\tilde{H}_b=k_x^2\hat{\sigma}_z+V(x)\hat{1}$. The eigenvalue equation
then decouples in two 1D Schr\"odinger equations $\mp d^2 \psi_\pm
/dx^2 =[E-V(x)]\psi_\pm$. As for a generic potential, all states of
the 1D Schr\"odinger equation are localized  \cite{MottTwose}, it
follows that the eigenstates of the ``1D bilayer'' hamiltonian are
also localized. This is the opposite conclusion to the ``1D
monolayer'' case. Here the conservation of pseudo-spin leads to
localization.

The conclusion that we draw on Klein tunneling across a step is
twofold: the transition in the step is possible if (1) there are
states available in the step at a matching energy (i.e. states of
negative kinetic energy) and (2) if the pseudo-spin conservation
permits such an inter-band transition. The latter provides a kind of
selection rule, reflecting whether the appropriate matrix element
for the inter-band transition vanishes or not. This matrix element
(squared) is usually called the chirality factor and is given by the
overlap of the incoming bispinor $(1,\sigma_x)/\sqrt{2}$ and the
transmitted bispinor $(1,\sigma_x')/\sqrt{2}$, which in 1D is
$(1+\sigma_x\sigma_x')/2=\delta_{\sigma_x,\sigma_x'}$. In the case
of the monolayer, the pseudo-spin is the direction of motion
$\sigma_x=E_\textrm{kin}/k_x$ and therefore the transition occurs
with unit probability. Whereas in the bilayer case, the pseudo-spin
is the band index $\sigma_x=\textrm{sign}(E_\textrm{kin})$, which
would obviously change in an inter-band transition
$\sigma_x'=-\sigma_x$, which is therefore strictly forbidden. These
results are reminiscent of the 2D case at normal incidence where
$T=1$ for the monolayer and $T=0$ for the bilayer \cite{Katsnelson},
and the corresponding inter-band chirality factors are $(1-\cos
\pi)/2=1$ and $(1-\cos(2\pi))/2=0$ respectively, see section
\ref{sharpstep}.

Note that at oblique incidence, the 2D massless case is quite unlike its 1D
counterpart. In fact, for a potential with $y$ translational invariance, there is an mapping
between the massless 2D case at oblique incidence ($k_y\neq 0$) and the 1D case of Dirac
electrons with a finite mass. Indeed, using the conservation of $k_y$,
the 2D eigenvalue equation of a massless Dirac electron
$(-i\hat{\sigma}_x \partial_x -i\hat{\sigma}_y \partial_y +V(x) \hat{1}) \psi(x,y)=E\psi(x,y)$ becomes a 1D equation for a massive Dirac electron:
\beq
(-i\hat{\sigma}_x \partial_x + m \hat{\sigma}_y +V(x) \hat{1}) \varphi(x)=E\varphi(x)
\eeq
where $\psi(x,y)=\varphi(x)\exp(ik_y y)$ defines the 1D wavefunction $\varphi(x)$ and $m\equiv k_y$ is the mass \footnote{The familiar form of the 1D Dirac equation is recovered by the unitary transformation $(\hat{\sigma}_x,\hat{\sigma}_y,\hat{\sigma}_z)\to (\hat{\sigma}_x,\hat{\sigma}_z,-\hat{\sigma}_y)$.}. This 1D equation is actually that originally considered by Klein with $V(x)=V_0 \Theta(x)$ \cite{klein} and by Sauter with $V(x)=Fx$ \cite{Sauter}.

\end{document}